\documentclass[a4paper,11pt]{article}
\pdfoutput=1 

\usepackage{jheppub} 

\usepackage[T1]{fontenc}
\usepackage[usenames,dvipsnames]{xcolor}

\def\be{\begin{equation}}
\def\ee{\end{equation}}
\def\bea{\begin{eqnarray}}
\def\eea{\end{eqnarray}}
\def\ba{\begin{array} }
\def\bac{\begin{array} {c}}
\def\bacc{\begin{array} {cc}}
\def\baccc{\begin{array} {ccc}}
\def\bacccc{\begin{array} {cccc}}
\def\ea{\end{array}}

\newcommand{\GL}{{\scriptscriptstyle\rm GL}}

\preprint{FTUAM-13-128, IFT-UAM/CSIC-13-016}

\title{\boldmath  Transitions in Dilaton Holography with Global or Local Symmetries}

\author[]{Alberto Salvio}

\affiliation[]{Departamento de F\'isica Te\'orica, Universidad Aut\'onoma de Madrid and \\ 
Instituto de F\'isica Te\'orica IFT-UAM/CSIC, Cantoblanco, 28049 Madrid, Spain}

\emailAdd{alberto.salvio@uam.es}

\abstract{We study various transitions in  dilaton holography, including those associated with the spontaneous breaking  of a global (superfluid case)  or local (superconductor case)  U(1) symmetry in diverse dimensions $d$. By analyzing the thermodynamics of the dilaton-gravity system we find that scale invariance is broken at low temperatures, as shown by   a nontrivial hyperscaling violation exponent in the infrared; increasing the temperature  we recover scale symmetry in a $d$ dependent way: while for $d=2+1$ a phase transition is found, for $d=3+1$ the transition is rather a  crossover. 
This is the expected behavior of QCD where the number of colors $N_c$ equals three (although in our holographic calculations $N_c\rightarrow \infty$).
 When the U(1) is preserved  and at low temperatures, the system is insulating  for arbitrary $d$ if the dilaton is appropriately coupled to the gauge field; for other couplings we also find a linear in temperature resistivity. We then determine the prediction of these models for  several quantities in the superconducting phase:  the DC and AC conductivity, the gap for charged excitations, the 
superfluid density, the vortex profiles, the coherence length, the  penetration depth and the  critical magnetic fields. We show that at low temperatures  some of these quantities differ qualitatively compared with the corresponding models without the dilaton, although the superconductor is robustly of Type II. The ratio of the gap over the critical temperature of the superconductor is studied in detail varying $d$ and  the couplings of the dilaton and then compared with the BCS value. A holographic renormalization is required in $d>2+1$ to compute some quantities (such as the AC conductivity and the penetration depth) and we explain in detail how to perform it. 

}
\begin{document} 

\maketitle
\flushbottom

\section{Introduction and summary of the article} \label{intro}
The gauge/gravity duality and its simplest realization, the anti de Sitter/conformal field theory (AdS/CFT) correspondence, have been extended in the last years to describe systems at finite temperature $T$ and density (associated with a U(1) symmetry). Much of the motivation for this setup comes from quantum chromodynamics (QCD) at finite temperature and baryon density, but more recent works have considered  applications to condensed matter  as well (for reviews see \cite{Hartnoll:2009sz,McGreevy:2009xe,Hartnoll:2009qx,Sachdev:2010ch,Pires:2010mt,Horowitz:2010nh,Hartnoll:2011fn}). The virtue of this approach, which we will take advantage of in this paper, is that it allows us to have information on a strongly coupled gauge theory
 (in the limit of a large number of gauge degrees of freedom) by doing {\it classical} calculations in a gravity theory. 

The minimal bulk field content  is given by the  $d+1$ dimensional metric and gauge field, which are dual to  the $d$ dimensional energy momentum tensor and  U(1) current. An interesting extension, however, is  provided by adding a real bulk scalar $\phi$ (the {\it dilaton}), which couples to the gauge field essentially by changing its  gauge coupling: $g\rightarrow g(\phi)$. Even without modifying the ultraviolet (UV) properties of the theory, the dilaton can modify the infrared (IR) limit relaxing the (often too strong) constraints that conformal invariance imposes:  a nontrivial (thus nonconstant) dilaton corresponds indeed to a nonvanishing beta function in the dual field theory. Most notably the presence of $\phi$ can allow a charged asymptotically AdS black hole (BH) to have vanishing entropy in the low temperature limit \cite{Gubser:2009qt,Goldstein:2009cv,Charmousis:2010zz}, contrary to what happens with the minimal field content. Dilatons, moreover, are typically present in the spectrum of string 
theories, which are natural UV completions for such effective holographic models.

In this paper we are interested in various transitions occurring  in dilaton holographic models in generic space-time dimension, including those that are associated with the spontaneous breaking of the U(1) symmetry. To this end we introduce a charged scalar dual to a condensing operator. If the U(1) current is identified with the electromagnetic one, such breaking corresponds to  superconductivity. The main motivation beyond holographic superconductors \cite{Hartnoll:2008vx,Horowitz:2008bn,Hartnoll:2008kx} is that they might lead to a better understanding  of high-$T$ superconductors, which certainly evade the weak coupling assumption of the theory proposed by Bardeen, Cooper, and Schrieffer (BCS). Cuprate high-$T$ superconductors are obtained by doping so called Mott insulators and in this process one typically observes an insulator/superconductor transition \cite{doping}. 
In ref. \cite{Salvio:2012at} we showed\footnote{Other studies of holographic superconductors with dilatons in $d=2+1$ can be found in \cite{Gouteraux:2012yr} (see also \cite{Gath:2012pg} for a previous study, which included the broken phase only).} for $d=2+1$ that an (approximate) insulator/superconductor transition can be obtained in dilatonic holographic superconductors for small enough $T$ and if $g(\phi)$ is big enough in the IR. Here we generalize the results of \cite{Salvio:2012at} to various space-time dimensions, including the physically interesting case $d=3+1$, and obtain the prediction of holography for important observables (such as the energy gap for charged excitations in the conductivity and the coherence length), which were not computed in \cite{Salvio:2012at}.

Another necessary ingredient for a realistic description of superconductivity through holography is the presence of a dynamical electromagnetic (EM) field. Dynamical gauge fields can be included by changing the usual Dirichlet AdS-boundary condition for the corresponding bulk field (suitable for {\it global} U(1) symmetries) with one of the Neumann type. Ref. \cite{Domenech:2010nf}  explained how to introduce a dynamical EM field in the original holographic superconductor model \cite{Hartnoll:2008vx,Horowitz:2008bn}. Here we make use of this method to define and study genuine superconductors in dilaton holography for various space-time dimensions. We also compare the resulting theory with the one with standard Dirichlet boundary conditions, which corresponds physically to a superfluid rather than to a superconductor. Superfluids indeed, in a sense that will be specified below, are equivalent to superconductors in the limit in which the EM field becomes nondynamical\footnote{Therefore, in order not 
to duplicate symbols  we will adopt the terminology of superconductivity rather than that of superfluidity, unless otherwise stated
(see however table \ref{table} for a dictionary between the two).}.

Let us give in detail the results of the present paper (and provide at the same time the organization). In the following section we define the holographic model and the gauge/gravity duality; we also recall the method of \cite{Domenech:2010nf} to introduce a dynamical gauge field. As we will see, for  $d>2+1$ this will force us to introduce a kinetic term localized on the AdS-boundary, which absorbs UV divergences through a holographic renormalization; more generally, such term is needed whenever the EM field has a nontrivial dependence on the space-time point, even if the EM field is nondynamical: for example this is the case when we determine the conductivity by computing the response of the system under a fixed (thus nondynamical) EM wave. In the same section we also study the solutions and the  thermodynamics of the dilaton-gravity system when the gauge field and the charged scalar are set to zero. We consider the general asymptotically AdS static BHs with $d-1$ rotation and translation symmetry, which 
were recently presented in \cite{Anabalon:2012ta,Acena:2012mr}. Fixed the parameters of the model there is only one solution with active dilaton (dilaton-BH henceforth), but the fact that the space is asymptotically AdS implies   the usual AdS Schwarzschild  black hole (S-BH) is always a solution. We therefore compare the free energies of the S-BH and dilaton-BH to determine the energetically favorable phase. We find that at low temperatures the system prefers the  dilaton-BH and in so doing breaks conformal invariance as shown by the emergence of a nontrivial hyperscaling violation exponent in the IR.  Increasing the temperature we observe a physically different behavior for $d=2+1$ and $d=3+1$: while in $d=2+1$ there is a phase transition and the system jumps  to the S-BH at some critical temperature, for $d=3+1$ we rather have a smooth crossover from the dilaton- to the S-BH. It is intriguing to notice that the confined/deconfined transition in
 real world ($N_c=3$) QCD is expected to be a crossover.  However, our results are in the limit of a large number of gauge degrees of freedom, which suggests that our dual theories cannot be identified with the large $N_c$ extrapolation of QCD.

In section \ref{Homogeneous superfluid transitions} we study the simplest example of holographic superfluid phase on top of the dilaton-BH: a homogeneous configuration with condensed charged operator at finite temperature and chemical potential $\mu$. As we will see a finite U(1) density ($\mu\neq 0$) is needed in order to have a regular solution of the field equations. This configuration always  exists when $0<T<T_c$, where $T_c$ is the critical temperature of the superconductor. Moreover, in this window of temperatures it is energetically favorable with respect to the normal phase (where the vacuum expectation value of the charged operator vanishes) like for the S-BH \cite{Hartnoll:2008vx,Horowitz:2008bn}. However, we observe that the dilaton has an impact on the condensate, which depends on its couplings to the other fields and on $d$.  

We then analyze the DC and AC conductivity on top of these BHs in section \ref{Conductivity}, considering both the normal and broken phase. While the presence of the horizon implies that the normal phase is conducting, if the coupling of the dilaton to the gauge field is appropriately chosen we have an (approximate) insulating behavior for the dilaton-BH. This effect is even stronger for small $T$ and  is enhanced in $d=3+1$ with respect to the $d=2+1$ case. For other choices of the dilaton couplings one can also realize a linear in temperature resistivity for small $T$.
In the broken phase we analyze the superfluid density $n_s$ and  study in detail  the gap of charge carriers $\omega_g$. In particular, we compute the ratio $\omega_g/T_c$ to compare it with the BCS value, 3.5. We find, like in the S-BH  \cite{Hartnoll:2008vx,Horowitz:2008bn}, a prediction which is robustly bigger than 3.5, but the dilaton does have an impact on such ratio. In particular,
when the couplings of the dilaton are such that the normal phase is (approximately) insulating $\omega_g/T_c$ is enhanced compared to the value found in the S-BH; there is also a range of parameters for which it is instead suppressed. As we have already pointed out, the calculation of the frequency dependent conductivity in $d>2+1$ requires a holographic renormalization. We describe in detail how to perform it and show that the DC conductivity, $n_s$ and $\omega_g$ are renormalization scheme independent. 

Finally, in section \ref{Inhomogeneous phases with broken U(1) symmetry} we study the response of the dilaton holographic superconductors under an external magnetic field $H$ and compare it with the case without the dilaton \cite{Hartnoll:2008kx,Montull:2009fe,Albash:2008eh,Nakano:2008xc,Maeda:2008ir,Ge:2010aa,Domenech:2010nf,Montull:2011im,Montull:2012fy,Salvio:2013ja}. On the one hand, the inclusion of magnetic fields is important as some of the most famous properties of superconductors emerge when $H\neq 0$. On the other hand, the inhomogeneous configuration which are triggered by $H$ allow us to illustrate the main differences between the global (superfluid) and local (superconductor) case. As we mentioned, in order to do this for $d>2+1$ we have to properly renormalize the EM field at least in the superconductor case. We find the Meissner effect and study the vortex phase including nontrivial couplings of the dilaton  to the gauge field and charged scalar. We obtain the coherence length $\xi'$,  
penetration depth $\lambda'$ and the critical magnetic fields $H_{c1}$ and $H_{c2}$. The latter observables allows us to show that  also these holographic superconductors are of Type II, both for $d=2+1$ and $d=3+1$, like those  studied in \cite{Domenech:2010nf,Montull:2012fy}. However, we find that $\lambda'$ and $H_{c1}$ have a qualitatively different $T\rightarrow 0$ limit with respect  to the S-BH \cite{Montull:2012fy}: $\lambda'$ may remain finite if the couplings of the dilaton are appropriately chosen both in $d=2+1$ and $d=3+1$ and $H_{c1}$ shows an analogue difference.  Interestingly, for these values of the couplings the normal phase is (approximately) insulating.

Section \ref{conclusions} contains some outlook for future work.

%

\section{The holographic model }

We consider the standard Einstein-dilaton action coupled to a U(1) gauge field $A_{\alpha}$ and a charged scalar $\Psi$ in $d+1$ space-time dimensions:
\be \hspace{-0.1cm}   S=\int d^{d+1}x\, \sqrt{-G}\left\{{{1\over16\pi
G_N}}\left[\mathcal{R}-(\partial_{\alpha} \phi)^2 - V(\phi) \right]\,-\frac{Z_{A}(\phi)}{4g^2}{\cal F}_{\alpha \beta}^2
-\frac{Z_{\psi}(\phi)}{L^2g^2}|D_\alpha\Psi|^2\right\}, \label{action} \ee 
where $\mathcal{R}$ is the Ricci scalar, $G_N$ is the Newton constant, $\phi$ is a real scalar (the dilaton)
and  we introduced ${\cal F}_{\alpha \beta}= \partial_\alpha A_\beta-\partial_\beta A_\alpha$  and $D_{\alpha}=\partial_{\alpha}-iA_{\alpha}$. A negative cosmological constant $\Lambda$ is included in $V(\phi)$, and the AdS radius is defined by $\Lambda=-d(d-1)/L^2$. We define $\phi$ such that $V(0)= \Lambda$. Also the  functions $Z_{A}(\phi)$ and  $Z_{\psi}(\phi)$ are generic and we choose the normalization of the matter fields in a way that  $Z_{A}(0)=Z_{\psi}(0)=1$. For the $Zs$ we only require to be regular and nonvanishing for any $\phi$ in order for the semiclassical approximation to be valid.  For simplicity,    we have not added any potential for $\Psi$.

In this article we focus on the most general static asymptotically AdS planar black hole with $d-1$ dimensional rotation and translation invariance, which has the following form:
\be ds^2=W(z)\left(-f(z)dt^2+dy^2+\frac{dz^2}{f(z)}\right),\qquad
 \phi=\phi(z)\ , \label{generalBH} \ee 
where $dy^2=\delta_{ij}dy^idy^j$ and $z$ is the holographic coordinate.  The AdS-boundary is located at $z=0$: 
we have $W(z)\simeq L^2/z^2$, $f(z)\simeq 1$ and $\phi(z)\simeq 0$ for $z\simeq 0$.

\subsection{The gauge/gravity correspondence}\label{The gauge/gravity correspondence}

Such $d+1$ dimensional classical gravitational theory  is supposed to be dual (through the gauge/gravity correspondence) to a $d$ dimensional 
strongly coupled CFT with a large number of gauge degrees of freedom. In the CFT side, the sources which generate the Green's functions are identified with the AdS-boundary values of bulk fields; the generating functional itself is conjectured
to be  the gravity action (\ref{action}) computed on a solution to the field equations with such AdS-boundary values 
as boundary conditions. For example,
\be \Psi|_{z=0}=\Psi_0\ , \qquad A_{\mu}|_{z=0}=a_{\mu}\ , \label{s-amu}\ee
can be identified with  the sources of a charged operator $\mathcal{O}$ and the current $J_\mu$ respectively. One then obtains the following vacuum expectation values:
\be 
\langle  J_\mu\rangle=\frac{L^{d-3}}{g^2}z^{3-d}{\cal F}_{z\mu}|_{z=0}\ ,  \qquad   \langle{\cal
O}\rangle=\frac{L^{d-3}}{g^2 }z^{1-d}D_z \Psi|_{z=0}\, .
\label{operator} \ee

In this formulation of the gauge/gravity  correspondence the boundary gauge field $a_\mu$ is  an external source for the  current and as such is a nondynamical field, introduced through a Dirichlet boundary condition (the second equation in (\ref{s-amu})). However, it is possible to make $a_{\mu}$ dynamical by substituting this condition with one of the Neumann type,
\be \frac{L^{d-3}}{g^2}z^{3-d}{\cal F}_z^{\,\,\, \mu} \Big|_{z=0} +\frac{1}{e_b^2}\partial_\nu {\cal F}^{\nu \mu}\Big|_{z=0}+J^\mu_{ext}=0\ ,
\label{maxwell2} \ee
where $J_{ext}^{\mu}$ is an external nondynamical current and we have added a bare boundary kinetic term $\frac{1}{e_b^2}\partial_\nu {\cal F}^{\nu \mu}\Big|_{z=0}$ for generality \cite{Domenech:2010nf}.
Condition (\ref{maxwell2}) requires to add  
\be
\int d^{4} x \left[
-\frac{1}{4e_b^2}{\cal F}_{\mu\nu}^2+A_\mu J_{ext}^\mu\right]_{z=0}
\label{extrat}
\ee
to  (\ref{action}). 
In the  case $d=2+1$ the bare boundary kinetic term is not necessary \cite{Witten:2003ya} and one can preserve conformal symmetry 
in the UV; in such case the dynamical $a_{\mu}$ is an emergent gauge field \cite{Domenech:2010nf}. 
When $d=3+1$ the situation is different. One way to see it is to notice that the current $\langle J_\mu \rangle$ contains a logarithmically divergent piece, 
\be \frac{1}{z}\partial_zA_\mu|_{z=0}=- \partial^\nu\mathcal{F}_{\nu\mu}\ln z|_{z=0}+ \,\,\, \mbox{finite terms}\ .\label{log-div}\ee
This divergence, however, can be reabsorbed in $e_b$:
\be \frac{1}{e_b^2}=\frac{1}{e_0^2}+\frac{L}{g^2}\ln z|_{z=0}+ \,\,\, \mbox{finite terms}\ ,\label{e0-def}\ee
where $e_0$ is the gauge coupling in the normal ($\Psi=0$) phase. In this case $a_\mu$ behaves very much as the four dimensional 
photon we are used to: the logarithmic divergence in (\ref{e0-def}) corresponds in the dual field theory to the logarithmic
divergence of the photon self-energy at very high momenta. For generic $d>3+1$ the situation is similar to the $d=3+1$ case, although the logarithmic divergence is substituted by a stronger one.

This method to render $a_\mu$ dynamical  has been originally applied to the holographic superconductor model \cite{Hartnoll:2008vx} (including diverse dimensions \cite{Horowitz:2008bn})
 in \cite{Domenech:2010nf} and we refer to this paper for further
details. Then, it has been applied to holographic p-wave \cite{Gubser:2008wv,Roberts:2008ns,Ammon:2008fc,Ammon:2009fe} and gapped \cite{Nishioka:2009zj,Horowitz:2010jq}
superconductors in 
\cite{Murray:2011gr,Gao:2012yw} and \cite{Montull:2012fy} respectively. In this paper we will 
use it to describe genuine superconductors in 
dilaton-gravity, extending the $d=2+1$ analysis of \cite{Salvio:2012at} to various dimensions and to compute additional observables.

\subsection{Solutions of the dilaton-gravity model}

We will work in the limit  $G_N\rightarrow0$
 such that  the effect of
$A_\alpha$ and $\Psi$ on $G_{\alpha\beta}$ and $\phi$ can be neglected.
In this limit the Einstein and dilaton equations are 
\be {\cal R}_{\alpha \beta}-\frac{G_{\alpha\beta}}{2}\left[\mathcal{R}-(\partial_{\gamma} \phi)^2 - V(\phi)\right] 
-\partial_\alpha \phi \ \partial_\beta\phi=0\label{Einstein-eqs}\ee
and 
\be \frac{2}{\sqrt{-G}}\partial_\alpha\left(\sqrt{-G}G^{\alpha\beta}\partial_\beta \phi\right)=\frac{\partial V}{\partial \phi}  \ee
respectively and the most general asymptotically AdS  black hole of the form (\ref{generalBH}) solving these equations   can be written as 
\bea \hspace{-2cm}W(z)&=&\frac{\nu^2\left(1+z/L\right)^{\nu-1}}{\left[\left(1+z/L\right)^{\nu}-1\right]^2}\ , \
\ \phi(z)=  \sqrt{\frac{\nu^2-1}{2}}\ln(1+z/L)\ , \nonumber \\ 
f(z)&=&1 + 3\left(\frac{L}{z_s}\right)^3\left\{\frac{\nu^2}{4-\nu^2}+(1+z/L)^2\left[1-\frac{(1+z/L)^{\nu}}{\nu+2}
-\frac{(1+z/L)^{-\nu}}{2-\nu}\right]\right\},\label{dilaton-BH-AdS4}
\eea 
for $d=3$ \cite{Anabalon:2012ta} , while, for $d=4$, we have \cite{Acena:2012mr} 
\bea W(z)&=&\frac{\nu^2\left(1+z/L\right)^{\nu-1}}{\left[\left(1+z/L\right)^{\nu}-1\right]^2}\ , \
\ \phi(z)=\sqrt{\frac{3(\nu^2-1)}{4}}\ln(1+z/L)\ , \nonumber \\
f(z)&=&1 -24\left(\frac{L}{z_s}\right)^4\left\{\frac{16}{(\nu^2-25)(9\nu^2-25)}
+\frac{(1+z/L)^{5/2}}{\nu^3}\left[\frac{(1+z/L)^{3\nu/2}}{3(3\nu+5)}\right.\right.\nonumber \\ 
& & \left. \left.+\frac{(1+z/L)^{-3\nu/2}}{3(3\nu-5)}-\frac{(1+z/L)^{\nu/2}}{\nu+5}-\frac{(1+z/L)^{-\nu/2}}{\nu-5}\right]\right\} ,
\label{dilaton-BH-AdS5}\eea 
where $\nu\geq 1$ is a dimensionless parameter and $z_s\geq 0$ has the dimension of a length.

 As usual we define the black hole horizon $z_h$ by $f(z_h)=0$. In order to study the theory at finite $T$, we perform the Euclidean continuation to imaginary time $it$ with period $\beta=1/T$. Then  $T=|f'(z_h)|/4\pi$, where the prime denotes the derivative with respect to $z$. 

In order for the functions $W(z)$, $f(z)$ and $\phi(z)$ to be solutions to the field equations the potential has to be appropriately chosen: we have an exponential form\footnote{Since the potential is even under $\phi\rightarrow -\phi$ we also have the solution in which we change the sign of $\phi$ in (\ref{dilaton-BH-AdS4}) and (\ref{dilaton-BH-AdS5}). However, without loss of generality we consider only the sign choice of (\ref{dilaton-BH-AdS4}) and (\ref{dilaton-BH-AdS5}).}
\be V(\phi)= \sum_{i=1}^{N_d} V_i^{(d)}e^{-\delta_i^{(d)} |\phi|} \ ,\ee
where $V_i^{(d)}$, $\delta_i^{(d)}$ and the natural number $N_d$ are constants.  
For $d=3$ we have $N_3=6$ and 
\bea & &\frac{z_s^3}{L}V^{(3)}_i=3  \frac{1-\nu }{2 + \nu} \ , \,\,\,\, 12\frac{1-\nu^2}{\nu^2-4}\ ,\,\,\,\, (1+\nu )\frac{ 3  \nu^2- \left(\nu^2-4\right)z_s^3/L^3}{\nu ^2(\nu -2)}\ , \nonumber \\ 
&&  3  \frac{\nu+1 }{2 - \nu}\ , \,\,\,\,
4(\nu^2-1)\frac{ 3  \nu^2- \left(\nu^2-4\right)z_s^3/L^3}{\nu ^2(\nu^2 -4) }\ ,  \,\,\,\,(\nu -1 )\frac{ 3  \nu^2- \left(\nu^2-4\right)z_s^3/L^3}{\nu ^2(\nu +2) }\ , \eea
where the entries on the right hand side correspond to $i=1,...,N_d$ respectively and 
\be  \sqrt{\frac{\nu^2-1}{2}}\,\,\delta^{(3)}_i=- \nu -1 \  , \,\,\,\, -1\ , \,\,\,\, 1-\nu  \ , \,\,\,\, \nu-1  \  , \,\,\,\, 1\ , \,\,\,\, \nu+1\ . \ee
For $d=4$ we have instead $N_4=7$,
 \bea  V^{(4)}_i&=& \frac{d_a(\nu+1)}{2(3\nu-5)}\ , \,\,\,\,\frac{d_a(\nu-1)}{2(3\nu+5)}\ , 
 \,\,\,\, \frac{5d_a(\nu^2-1)}{9\nu^2-25}\ , \,\,\,\, \frac{d_b}{3\nu+5}\ , \,\,\,\, \frac{d_b}{3\nu-5}\ ,
 \,\,\,\,  \nonumber \\ && \frac{d_c(\nu-1)}{(\nu+5)(3\nu+5)}\ , \,\,\,\,  \frac{d_c(\nu+1)}{(\nu-5)(3\nu-5)}\ ,
\nonumber\eea
where
\be d_a= \frac{9\nu^2-25}{\nu^2L^2}\left[-\frac{3}{2}+\frac{4(12L^2/z_s^2)^2}{(\nu^2 -25)(9\nu^2-25)}\right]\ , 
\,\,\,\, d_b=\frac{(12 L)^2}{\nu^3 z_s^4}\frac{5(1-\nu^2)}{\nu^2-25}\ , \,\,\,\, d_c=-\frac{(12 L)^2}{3\nu^3 z_s^4} \nonumber
\ee
and 
\bea \sqrt{\frac{3(\nu^2-1)}{4}}\,\, \delta^{(4)}_i&=& 1-\nu  \  , \,\,\,\, \nu +1 \ , \,\,\,\, 1\  ,\nonumber  \\  \,\,\,\, & & \,\,\,\,-(3+\nu)/2\ , \,\,\,\, 
(\nu-3)/2\ , \,\,\,\,- 3(\nu+1)/2\ , \,\,\,\, 3(\nu-1)/2\ .\eea

We see that for $\nu=1$ we recover the S-BH: $W(z)=L^2/z^2$, $f(z)=1-(z/z_s)^d$ and $\phi=0$. Therefore 
$\nu-1$ measures how strongly scale invariance is broken in the IR. Indeed, if we take the zero temperature limit we find  $f(z)\rightarrow 1$ and 
$W(z)$ in the deep IR, $z\gg L$, can be approximated by $W(z)\simeq \nu^2(L/z)^{\nu+1}$, showing that $\nu-1\neq0$ corresponds to 
a nontrivial hyperscaling violation exponent $\theta$ \cite{Fisher:1986zz,Huijse,Dong:2012se}. The explicit expression of $\theta$ is 
\be \theta =\frac{1-\nu}{2}(d-1)\ ,\ee 
showing that $\theta$ is $d$-dependent and always $\theta \leq 0$. Hyperscaling violation exponents with these properties can be realized in
string theory \cite{Dong:2012se}.

\section{Transitions in dilaton-gravity}

 Since  $V(0)=\Lambda<0$, 
in addition to the solutions with nonvanishing dilaton in eqs. (\ref{dilaton-BH-AdS4}) and (\ref{dilaton-BH-AdS5}),
we always have the usual S-BH, independently on the value of $\nu$. 
In this section we determine which solution among the S-BH and the dilaton-BH is energetically
favorable.

In order to do so we have to study the  free energy $F$ of the system, which is given by $F=TS_E$, 
where $S_E$ is the on-shell Euclidean action. By using (\ref{action}) we have
\be F=\frac{1}{16\pi G_N}\int d^dx \sqrt{-g}\left[-{\cal R}+\left(\partial_\alpha \phi\right)^2+V(\phi)\right]\ , \label{free-energy}\ee
where the integral is performed over the spatial dimensions. By taking the trace of the Einstein equations in (\ref{Einstein-eqs}), we obtain
an expression for ${\cal R}$, which inserted in (\ref{free-energy}) gives 
\be F=\frac{1}{16\pi G_N}\frac{2}{1-d}\int d^dx \sqrt{-g}\ V(\phi)\ .\ee

\begin{figure}[ht]
   {\hspace{-1.5cm}}
  \hspace{4cm} \includegraphics[scale=0.7]{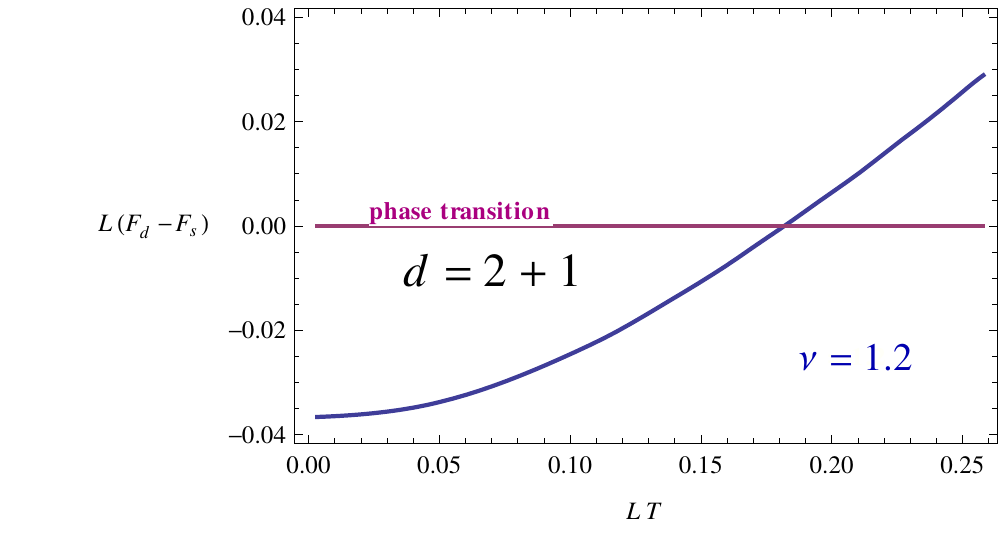}  
 
   \caption{{\small The difference between the  dilaton-BH and the S-BH free energy densities 
   (up to the gravitational constant $1/16\pi G_N$) vs. the temperature for $d=2+1$.}}
\label{phase-transition-D-d-BH}
\end{figure}

By substituting the ansatz in (\ref{generalBH}) into this equation we then obtain
\be F=\frac{V_{d-1}}{16\pi G_N}\int_\epsilon^{z_h} dz \frac{2}{1-d} W^{\frac{d+1}{2}}(z) \ V(\phi(z))\ , \label{Fansatz} \ee
where $V_{d-1}$ is the volume of the $d-1$ dimensional space of the CFT and the parameter $\epsilon$ represents a
UV cut-off. Let us start with the $d=3$ case for which the integrand in (\ref{Fansatz})
is $-W^2(z)V(\phi(z))$ and close to the AdS-boundary we have
\be \left\{ \bac - W^2(z)V(\phi(z))\simeq \frac{6L^2}{z^4} +\frac{\nu^4+19-20\nu^2}{120L^2} 
+ \mathcal{O}(z)\qquad \mbox{($d=3$ dilaton-BH)}\ ,\\ \\
-W^2(z)V(\phi(z))\simeq \frac{6L^2}{z^4} \qquad \mbox{($d=3$ S-BH)}\ .
\ea \right. \nonumber\ee
From this equation we see not only that the UV asymptotics of the two solutions are the same but also that the UV divergences cancel in the 
difference $F_d-F_s$, where $F_d$ and $F_s$ are the free energy densities for the dilaton-BH and the S-BH respectively. Performing explicitly 
the integral in (\ref{Fansatz}), we find that the S-BH is energetically favorable at high temperatures while, decreasing the temperature below 
a critical threshold, 
a phase transition occurs such that the system jumps to the dilaton-BH. This situation is described in figure \ref{phase-transition-D-d-BH}.

When $d=4$ we have a qualitatively different physics. The integrand in (\ref{Fansatz})
is $-2W^{5/2}(z)V(\phi(z))/3$ and close to the AdS-boundary we have
\be \left\{ \bac -\frac{2}{3} W^{5/2}(z)V(\phi(z))\simeq \frac{8L^3}{z^5} +\frac{L(1-\nu^2)}{6z^3} +\frac{(1-\nu^2)(17\nu^2-113)}{720L^2}
+ \mathcal{O}(z)\qquad \mbox{($d=4$ dilaton-BH)}\ ,\\ \\
-\frac{2}{3}W^{5/2}(z)V(\phi(z))= \frac{8L^3}{z^5} \qquad \mbox{($d=4$ S-BH)}\ .
\ea \right. \nonumber \ee
Although the dilaton- and S-BH are both asymptotically AdS and as such their free energy densities have the same leading UV divergence, 
the difference $F_d-F_s$ is not UV finite for $d=4$, but is dominated by the next-to-leading divergence of the dilaton
free energy, $~(1-\nu^2)/\epsilon^2$.  Since $1-\nu^2\leq 0$ for the dilaton-BH, it follows that the solution with active dilaton 
is always energetically favorable with respect to the S-BH. Nevertheless, when the temperature is increased, $z_h$ becomes small in units 
of $L$ and the dilaton-BH resembles more and more the S-BH. In other words we observe a crossover from a situation where scale invariance 
is broken to one where scale invariance is gradually recovered. The latter region of the phase diagram can be interpreted as a deconfined phase.
The fact that here we have  a crossover rather than a phase transition is intriguing as deconfinement is expected to correspond to a crossover in the phase
diagram of real world QCD (with the number of colors $N_c=3$) \cite{Pawlowski:2010ht}. However, we should point out that increasing $N_c$ in QCD one observes a stronger deconfining phase transition, instead of a crossover \cite{Lucini:2002ku, Lucini:2003zr, Lucini:2005vg}. At the very least this suggests that the dual of the holographic models we are studying is not the large $N_c$ extrapolation of QCD, but of distinct field theories. 

\section{Homogeneous superfluid transitions} \label{Homogeneous superfluid transitions}

Let us now look for the simplest superconducting  ($\Psi\neq 0$)  solutions and study how and when phase transitions from the normal phase to such superconducting solution occur.  

As we will see, there is no regular solution with $\Psi\neq 0$ if the gauge field is set to zero and a way to avoid this problem is to have $A_0\neq 0$. Therefore, we take the ansatz
\be \Psi=\psi(z)\ ,\qquad A_0=A_0(z)\ ,\label{ansatz} \ee
where $\psi$ is a real function, and the spatial components of $A_{\alpha}$  are set  to zero. 
The requirement that $A_\alpha A^\alpha$ is finite implies $A_0|_{z=z_h}=0$ (see ref. \cite{Hartnoll:2008vx}). Thus, in order to keep $A_0$ nonvanishing somewhere we impose $A_0\neq 0$ at the AdS-boundary; remembering also that a superconducting solution breaks the U(1) spontaneously rather than explicitly, we impose
\be \Psi_0=0\ , \qquad a_0=\mu\ , \label{UV}\ee
where $\mu$ is a nonvanishing chemical potential. So the system is at finite density. 

The ansatz in (\ref{ansatz})
corresponds to the field equations
\bea \frac{\partial_z\left(W^{(d-1)/2} f Z_{\psi}(\phi) \partial_z\psi\right)}{W^{(d-1)/2}Z_{\psi}(\phi)} +\frac{A_0^2 }{f} \psi =0\ ,
\, \,\, \frac{\partial_z \left(W^{(d-3)/2}Z_A(\phi)\partial_zA_0\right)}{W^{(d-3)/2}Z_\psi(\phi)}-\frac{2W \psi^2}{L^2f}  A_0=0 \label{field-equations}\eea  
and here we can see that there are no regular superconducting solutions with $A_\alpha=0$; indeed setting
$A_0=0$ in the equations above we find $\partial_z \psi\propto \left(W^{(d-1)/2} f Z_{\psi}(\phi) \right)^{-1}$, which is singular at the black hole horizon. 

The boundary conditions in (\ref{UV}) are not sufficient to  determine the solution to eqs. (\ref{field-equations}). However, the regularity at the black hole horizon provides us with other two conditions. Indeed, from the second equation in (\ref{field-equations}) we see that in order for $A_0$ (and its first and second derivatives) to be regular at $z=z_h$ we should impose $A_0=0$, as previously found by demanding $A_\alpha A^\alpha$ to be regular. More precisely we see that $A_0\rightarrow 0$ as $z\rightarrow z_h$ at least as fast as $f$. Then, we can see that for $z\neq z_h$ the first equation in (\ref{field-equations}) reduces to $W^{(d-1)/2}f' Z_{\psi}(\phi) \partial_z\psi=0$ with corrections which go to zero as $z\rightarrow z_h$. Since $W$, $f'$ and $Z_\psi(\phi)$ are also  regular at the black hole horizon, we find 
\be \partial_z\psi =0\ , \qquad  A_0=0\qquad (\mbox{at}\, \, z=z_h)\ .  \label{IR}\ee

It is worth noting that while (\ref{UV}) represent external conditions and as such are chosen according to the physics we want to describe, (\ref{IR}) are rather fixed conditions (as required by regularity). From the dual CFT perspective, $\Psi_0$ and $\mu$ are external sources while the 
conditions in  (\ref{IR}) (and  eqs (\ref{field-equations})) correspond to the dynamical response of the CFT.

\begin{figure}[t]
   \begin{tabular}{cc}
   {\hspace{-0.3cm}}
   \includegraphics[scale=0.65]{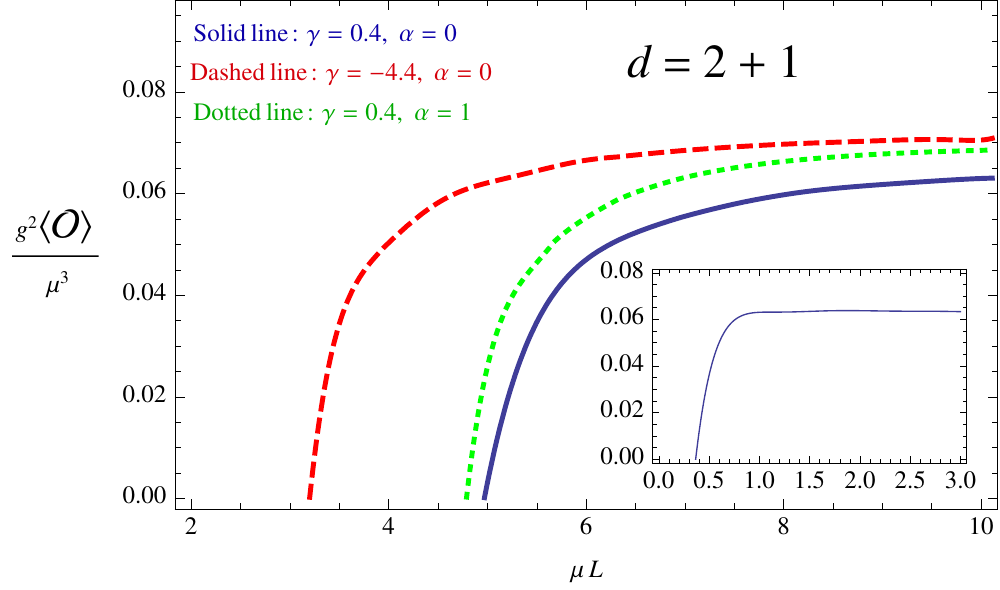}  
 
     {\hspace{1.cm}} 
    \includegraphics[scale=0.65]{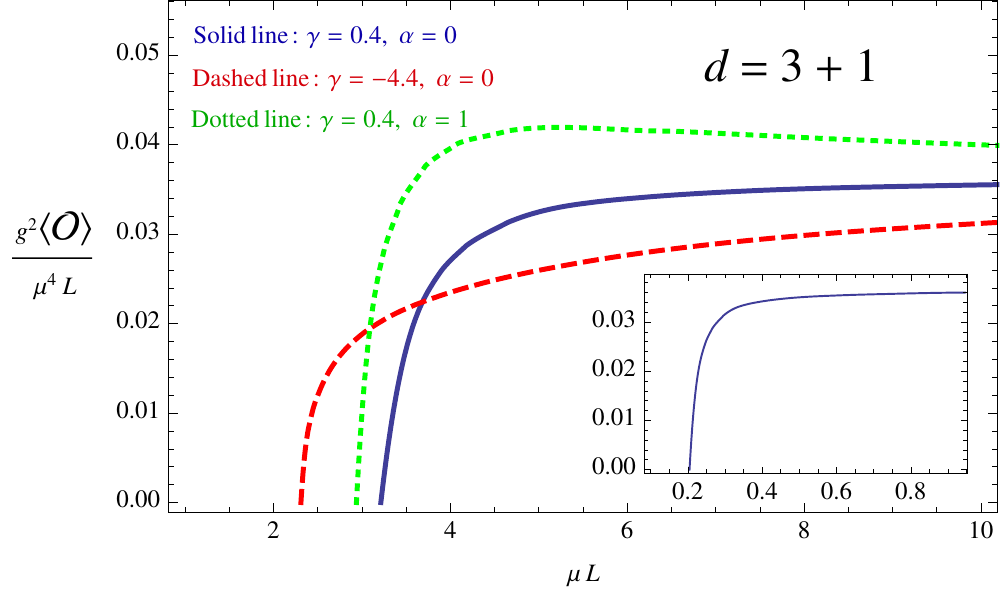}  
    \end{tabular}
   \caption{\label{condensate-vs-muL}The condensate as a function of  $\mu L$. We set $\nu=1.2$, $Z_A(\phi)=e^{\gamma\phi}$ and $Z_\psi=e^{\alpha \phi}$.
   We choose  $T\simeq 0.010/L$ in the inset and $T\simeq 0.15/L$ everywhere else. }

\end{figure} 

We have solved eqs. (\ref{field-equations}) with the boundary conditions  (\ref{UV}) and (\ref{IR}) 
and we plot the resulting $\langle \mathcal{O} \rangle$ as a function of $\mu L$ in figure \ref{condensate-vs-muL}. 
We see that, like in the holographic models without the dilaton \cite{Hartnoll:2008vx, Horowitz:2008bn}, 
there is a critical value of $\mu L$ above which the condensate becomes nonzero. 
Such critical value, however,  is a function not only of $T/L$, but also of $\nu$, $Z_A$ and $Z_\psi$ and may be inverted to have $T_c/L$
in terms of the chemical potential, $\nu$ and the dilaton couplings to $A_\alpha$ and $\Psi$.  Moreover, we have checked that, when 
the superconducting solution exists, it is energetically favorable with respect to the normal phase, as it can be shown by comparing their free energies.
From figure  \ref{condensate-vs-muL} we see that the effect of the dilaton functions couplings is substantially different in the $d=2+1$ 
and $d=3+1$.

\section{Conductivity} \label{Conductivity}

We now turn to the (complex) conductivity $\sigma$. We will  study this quantity both in the normal and superconducting phases. The conductivity in the normal phase will be given explicitly in section \ref{conductivity-unbroken}; this will allow us, among other things, to
 investigate the nature of the superfluid phase transition studied in the previous section.  The conductivity in the broken phase will then be presented in section \ref{conductivity-broken}; from this quantity one can indeed observe an infinite DC conductivity and that the system develops a gap for charged carriers.

Consider a small time-dependent gauge field along a spatial dimension $x$,
\be A_x (t,z)=\mathcal{A}(z)e^{i\omega (p(z)-t)}\ , \label{wave-decomposition}\ee
 which is induced by a small EM field at the
 AdS-boundary, $a_x (t)$. Here $\mathcal{A}$ and $p$ are real functions of $z$. The system responds creating a current linear in $a_x$: $\langle J_x\rangle= \sigma E_x$, where  $E_x=-\partial_t a_x$ is the electric field. $\sigma$ can be computed once the equations of the other fields (other than the EM one) are solved on the EM background, $a_x(t)$; therefore, the conductivity does not depend on the dynamics of the  EM field. 
 
 The current $\langle J_x\rangle$ can be computed with the gauge/gravity correspondence (the first equation in (\ref{operator})), to obtain 
 \be \langle J_x\rangle=\frac{L^{d-3}}{g^2}\left[z^{3-d}\left(A_x \frac{\mathcal{A}'}{\mathcal{A}}+i\omega A_xp'\right)\right]_{z=0}, \label{current-conductivity}\ee
 having chosen a gauge such that $A_z=0$.
 Notice that this expression is generically UV divergent when $d>2+1$ because the EM field we are considering is time-dependent: 
 by looking at eq. (\ref{log-div}), which holds for $d=3+1$, we see that  $\langle J_\mu\rangle$ 
 is logarithmically divergent whenever $\partial^\nu\mathcal{F}_{\nu\mu}\neq 0$; that singularity is replaced
 by an even stronger one when $d>3+1$. 
 We remove this divergence by adding a boundary counter term of the form given in the first term of (\ref{extrat}).
  In the particular case $d=3+1$ this means that $1/e_b$ is logarithmically divergent in order for $e_0$ (defined in (\ref{e0-def})) to be finite.
  This is nothing but a renormalization procedure and make expressions like (\ref{current-conductivity}) finite. In the case of 
  $\langle J_x\rangle$ we have
 \be \langle J_x\rangle\rightarrow \left[\frac{L^{d-3}}{g^2}z^{3-d}\left(A_x \frac{\mathcal{A}'}{\mathcal{A}}+i\omega A_xp'\right)+
 \frac{\omega^2}{e_b^2}A_x\right]_{z=0} . \label{current2-conductivity}\ee
 %
%
  By substituting eq. (\ref{current2-conductivity}) in the definition of the conductivity $\sigma = \langle J_x\rangle/E_x$ we obtain 
  \be \sigma=\left[\frac{L^{d-3}}{g^2}z^{3-d}\left(p'-\frac{i\mathcal{A}'}{\omega \mathcal{A}} \right)-i\frac{\omega}{e_b^2}\right]_{z=0} \ .
  \label{sigma}\ee 
  So the real and imaginary parts of the conductivity are given by
  \be \mbox{Re}[\sigma] =\frac{L^{d-3}}{g^2}\left(z^{3-d}\ p'\right)_{z=0}, \qquad 
  \mbox{Im}[\sigma]=-\left(\frac{L^{d-3}}{g^2\omega}z^{3-d}\ \frac{\mathcal{A}'}{ \mathcal{A}}+\frac{\omega}{e_b^2} \right)_{z=0}.  \label{ReImsigma}\ee
    From this result we see that the boundary kinetic term only changes Im$[\sigma]$ and it does so only by adding a term proportional 
    to $\omega$. This implies that physical quantities such as the DC conductivity, the superfluid density $n_s$ and the gap of charge carriers (in the 
    superfluid phase) are independent on how we choose  $1/e_b$, namely they are independent on the particular renormalization scheme.

    Since $a_x(t)$ is small we can determine all the quantities needed to compute $\sigma$ in (\ref{sigma}) by solving the linearized Maxwell equation
\be \frac{\partial_z\left(W^{(d-3)/2}f Z_A(\phi)\partial_z A_x\right)}{W^{(d-3)/2}f Z_A(\phi)}+ \frac{ \omega^2}{f^2} A_x-2\frac{WZ_\psi(\phi)\psi^2}{L^2Z_A(\phi)f}A_x= 0\ . \label{linearizedEOM}\ee 
In the last term of the equation above we are taking into account the possibility that the system is  superconducting, $\psi\neq 0$. Eq. (\ref{linearizedEOM}) has to be solved with appropriate boundary conditions. At the black hole horizon we impose,  as required by regularity \cite{Hartnoll:2009sz}, that the EM wave is going towards the horizon (ingoing boundary condition), that is
\be p'(z)f(z)\rightarrow 1 \qquad \mbox{as} \quad z\rightarrow z_h\ . \label{IRBC}\ee 

   Inserting (\ref{wave-decomposition}) into (\ref{linearizedEOM}) (and after that dividing by $A_x$) we obtain an equation whose imaginary part can be solved analytically, to obtain
\be p'(z)=g^2\mbox{Re}[\sigma] \ \frac{\left[\left(z/L\right)^{d-3}W^{(d-3)/2}\right]_{z=0}}{W^{(d-3)/2}(z)Z_A(\phi(z))f(z)} \left(\frac{\mathcal{A}(0)}{\mathcal{A}(z)}\right)^2, \label{Imeq}\ee 
where we used the expression for Re$[\sigma]$ given in (\ref{ReImsigma}).
A consequence is that Re$[\sigma]$ cannot be zero without violating eq. (\ref{IRBC}).
Therefore, the black hole horizon generates an imaginary part of the amplitude $\mathcal{A}(z)e^{i\omega p(z)}$, which in turn
 results in a nonvanishing Re$[\sigma]$. 
 In the fluid mechanical interpretation Re$[\sigma]\neq 0$ corresponds to a fluid, as opposed 
to a solid which has a vanishing Re$[\sigma]$ (see \cite{Montull:2012fy}).  Inserting (\ref{Imeq}) into  (\ref{linearizedEOM}) we obtain an equation for $\mathcal{A}$, 
\be  \frac{\left(W^{(d-3)/2}f Z_A(\phi)s'\right)'}{W^{(d-3)/2}f Z_A(\phi)}+(s')^2+\frac{\omega^2}{f^2}\left(1-\frac{e^{-4s}}{W^{d-3}Z_A^2}\right)-2\frac{WZ_\psi(\phi)\psi^2}{L^2Z_A(\phi)f}=0\ , \label{s-equation}\ee 
where $$s(z)\equiv \ln\frac{\mathcal{A}(z)}{\mathcal{A}(0)}-\frac{1}{2}\ln\left( p'W^{(d-3)/2}\right)_{z=0},$$ which, once inserted in (\ref{Imeq}), gives an expression of $p'$ in terms of $s$,
\be p'=\frac{e^{-2s}}{W^{(d-3)/2}Z_A(\phi) f}\ . \label{p-in-terms-of-s}\ee
We solve eq. (\ref{s-equation}) with the regularity boundary conditions (at $z=z_h$)
\be \hspace{-0.2cm}s=-\frac{1}{2}\ln\left(W^{(d-3)/2} Z_A\right), \qquad s'=\frac{2f'W Z_\psi \psi^2 /L^2 -\omega^2 \left[2Z_A'+(d-3)Z_AW'/W \right]}{Z_A(4\omega^2+f'^2)}\ .  \label{bc-s}\ee
The first of these conditions comes from the requirement that the EM wave must go towards the horizon, eq. (\ref{IRBC}), and eq. (\ref{p-in-terms-of-s}); it ensures that the terms going like $1/(z-z_h)^2$ in eq. (\ref{s-equation}) vanish. The second one ensures that the next to leading terms, $\sim 1/(z-z_h)$, vanish as well. 
From the solutions to (\ref{s-equation}) and (\ref{bc-s}) we can compute  $\sigma$ with 
  \be \mbox{Re}[\sigma] =\frac{1}{g^2}\left(\frac{L^{d-3}\ e^{-2s}}{z^{d-3}W^{(d-3)/2} }\right)_{z=0}, 
  \qquad \mbox{Im}[\sigma]=-\left[\frac{1}{g^2\omega}\left(\frac{z}{ L}\right)^{3-d} s'+\frac{\omega}{e_b^2}\right]_{z=0}.  \label{ReImsigma2}\ee

    Before presenting $\sigma$ explicitly let us observe that the behavior of Re$[\sigma]$ at high frequencies\footnote{Im$[\sigma]$ 
    at high frequencies  is instead renormalization scheme dependent as different choices for the finite part of $1/e_b$
can change the term proportional to $\omega$. When $d<3+1$ and the renormalization is not required,  taking $e_b\rightarrow \infty$ (that
is without putting the boundary kinetic term) one obtains Im$[\sigma]\rightarrow 0$.} can be easily determined. Indeed, in this limit (\ref{s-equation}) and (\ref{bc-s})  tell us
    \be s= -\frac{1}{4}\ln\left(W^{d-3}Z_A^2\right) +O(1/\omega^2)\ee 
so one easily finds Re$[\sigma]\rightarrow 1/g^2$ when $d=2+1$ and Re$[\sigma]\rightarrow +\infty$ for 
$d>2+1$.  This result is in agreement with \cite{Horowitz:2008bn}, where they analyzed $\sigma$ in the superconductive case (and in the absence of the dilaton). These findings, however, do not rely on the fact that the system is superconducting and hold in the normal phase as well: as expected the system does not know if the symmetry is broken at high enough energies.

\begin{figure}[t]

   \begin{tabular}{cc}
   {\hspace{-0.7cm}}
   \includegraphics[scale=0.71]{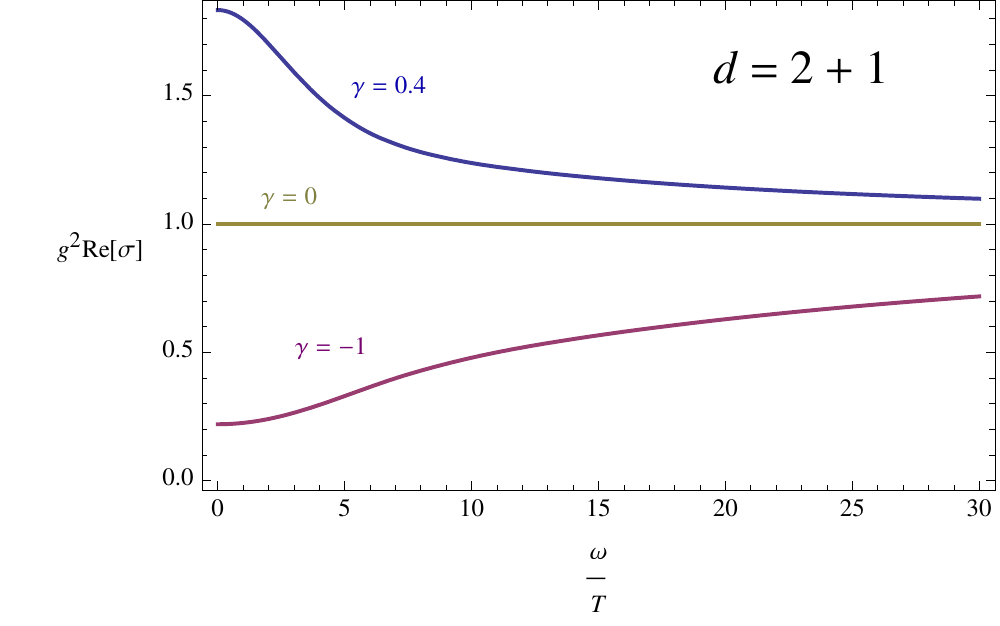}  
      {\hspace{0.2cm}} 
    \includegraphics[scale=0.73]{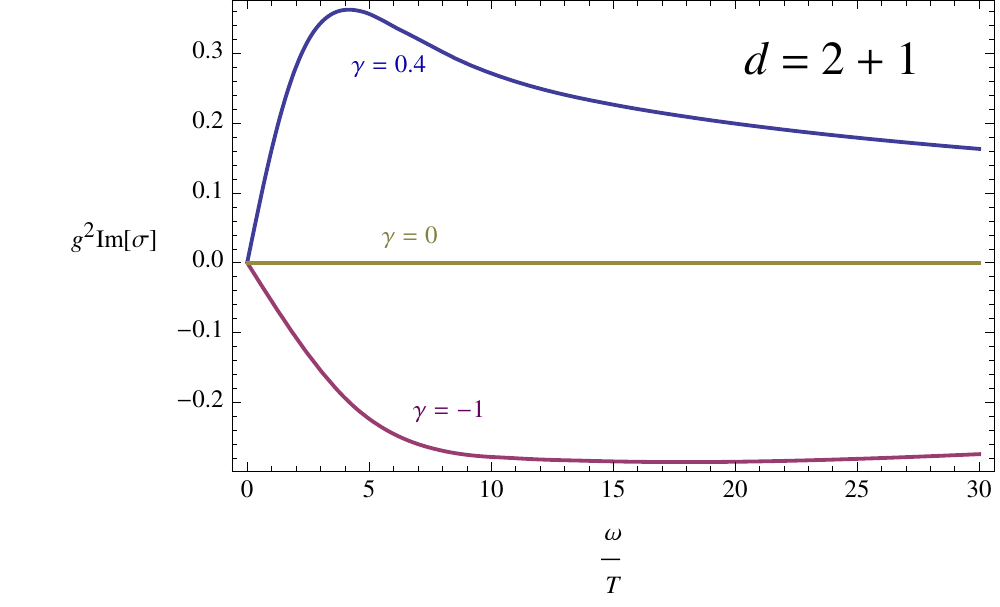}  
    \end{tabular}
    
    \vspace{0.4cm}
    
    \begin{tabular}{cc}
   {\hspace{-0.3cm}}
   \includegraphics[scale=0.67]{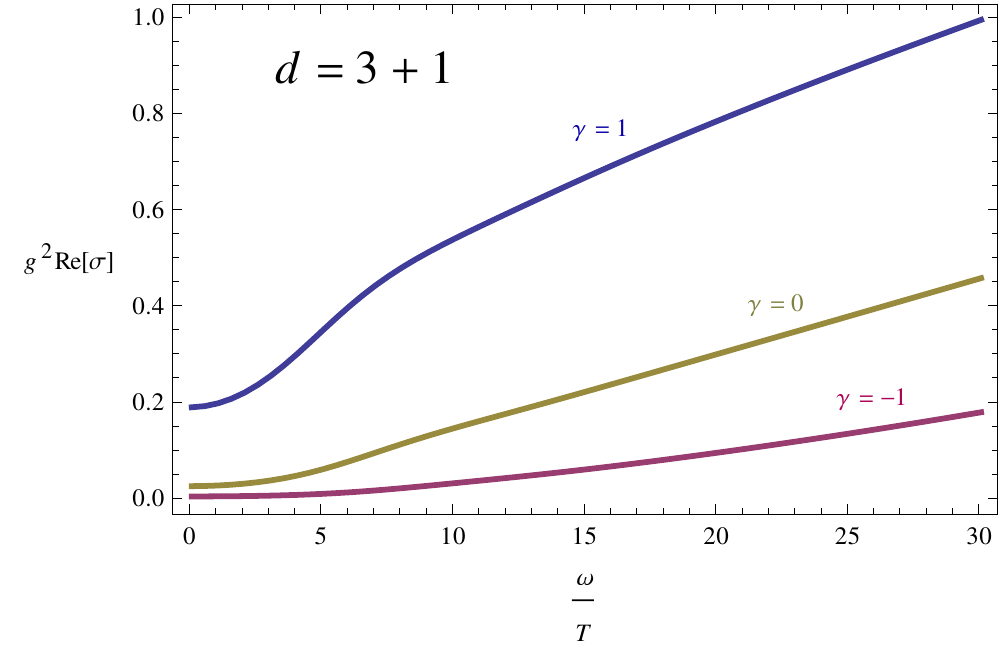}  
      {\hspace{0.8cm}} 
    \includegraphics[scale=0.69]{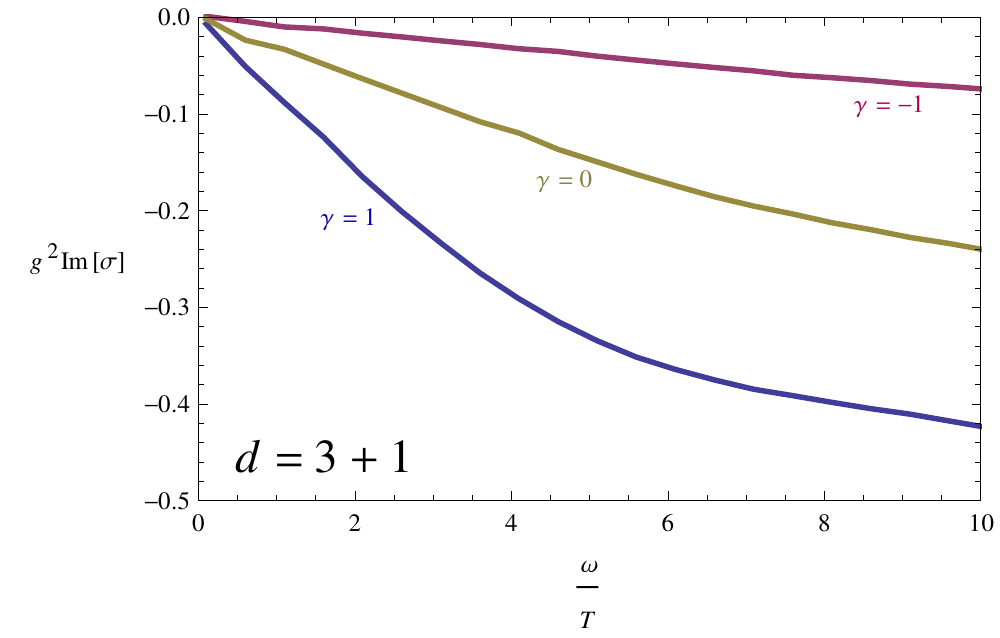}  
    \end{tabular}
    
   \caption{\label{sigma-NP}The conductivity as a function of the frequency in the unbroken phase, $\langle \mathcal{O }\rangle=0$.
   We set $T\simeq 0.010/L$, $\nu=1.2$, $Z_A(\phi)=e^{\gamma\phi}$ and $Z_\psi=1$. Moreover, we choose  $g^2/e_b^2=L$ ln$(z/L)|_{z=0}$
for $d=3+1$,  
   while $e_b/g\rightarrow \infty$ for $d=2+1$.}

\end{figure}

\subsection{Conductivity in the unbroken phase}\label{conductivity-unbroken}
Let us now focus on the normal phase (see also the related study \cite{Alishahiha:2012ad} for a charged BH). A particularly interesting case is the low frequency limit $\omega\rightarrow 0$ as it allows us, in particular, to obtain the DC conductivity. We find
\be \lim _{\omega \rightarrow 0} \mbox{Re}[\sigma]=\frac{1}{g^2}\left[W^{(d-3)/2} Z_A\right]_{z=z_h}, \qquad  \lim _{\omega \rightarrow 0}\mbox{Im}[\sigma]=0. \label{DCsigma}\ee
The result for the DC conductivity in (\ref{DCsigma}) is interesting as it tells us that this quantity is suppressed when $Z_A$ is small at the black hole horizon, resembling an insulator, while it is enhanced if $Z_A|_{z=z_h}$
is large.

These effects become strong in the low temperature region where the dilaton is logarithmically large, eqs. (\ref{dilaton-BH-AdS4}) and (\ref{dilaton-BH-AdS5}). Indeed, taking the typical exponential form $Z_A(\phi) = e^{\gamma \phi}$, where $\gamma$ is a constant, we find {\it at low temperatures}
\be \lim _{\omega \rightarrow 0} \mbox{Re}[\sigma] \sim T^{-\gamma \sqrt{(\nu^2-1)/2}} \quad (d=2+1)\ , \quad \lim _{\omega \rightarrow 0} \mbox{Re}[\sigma] \sim T^{(1+\nu)/2-\gamma \sqrt{3(\nu^2-1)/4}}\quad  (d=3+1)\ . \label{sigma-vs-T} \nonumber \ee
Therefore, taking $\gamma <0$ (that is suppressing  $Z_A|_{z=z_h}$), we have an almost insulating material. The $d=2+1$ result reproduces that found in \cite{Salvio:2012at}. Here we observe that when $d=3+1$ this effect is even stronger (due to the contribution of  $(1+\nu)/2$) and in order to obtain a low temperature insulator it is sufficient to have $\gamma<\sqrt{(1+\nu)^2/[3(\nu^2-1)]}$. It is interesting to note that, when the normal phase is approximately insulating, the superfluid phase transitions studied in section \ref{Homogeneous superfluid transitions} are insulator/superconductor transitions. These are typically observed in the low temperature phase diagram of cuprate high-$T$ superconductors. At the same time we observe that when $\gamma=\sqrt{2/(\nu^2-1)}$ (for $d=2+1$) and $\gamma=(3+\nu)/\sqrt{3(\nu^2-1)}$ (for $d=3+1$) we have a linear in temperature resistivity, a phenomenon which is found in many unconventional superconductors\footnote{See also \cite{Donos:2012ra} for another mechanism to 
obtain a linear in temperature resistivity in holography. }.

In figure \ref{sigma-NP} we give the conductivity as a function of $\omega/T$. 
That figure shows that  Re$[\sigma(\omega)]$ is large (small) when $Z_A(\phi)$ is large (small) at the horizon; 
this behavior was expected because we showed above  that it occurs at least for $\omega \rightarrow 0$.

\begin{figure}[h]

   \begin{tabular}{cc}
   {\hspace{-0.7cm}}
   \includegraphics[scale=0.64]{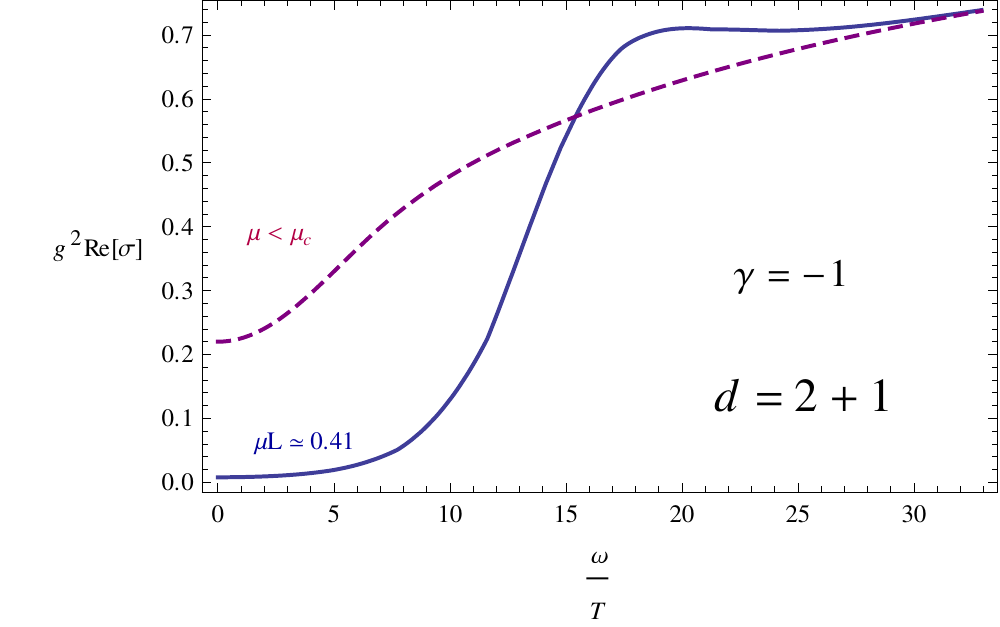}  
      {\hspace{0.2cm}} 
    \includegraphics[scale=0.66]{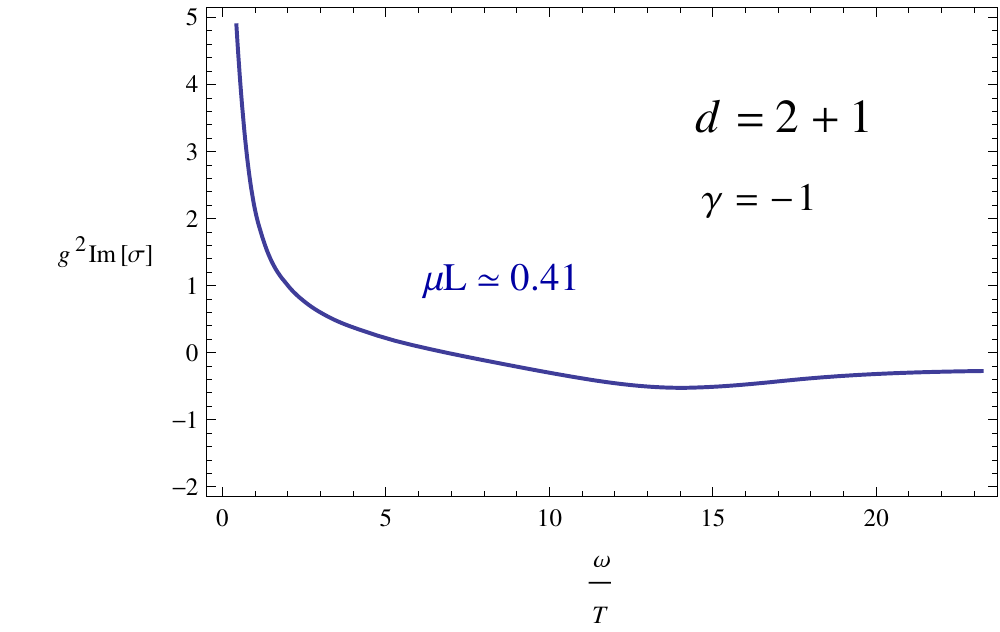}  
    \end{tabular}
    
     \vspace{0.4cm}
    \begin{tabular}{cc}
   {\hspace{-0.7cm}}
   \includegraphics[scale=0.65]{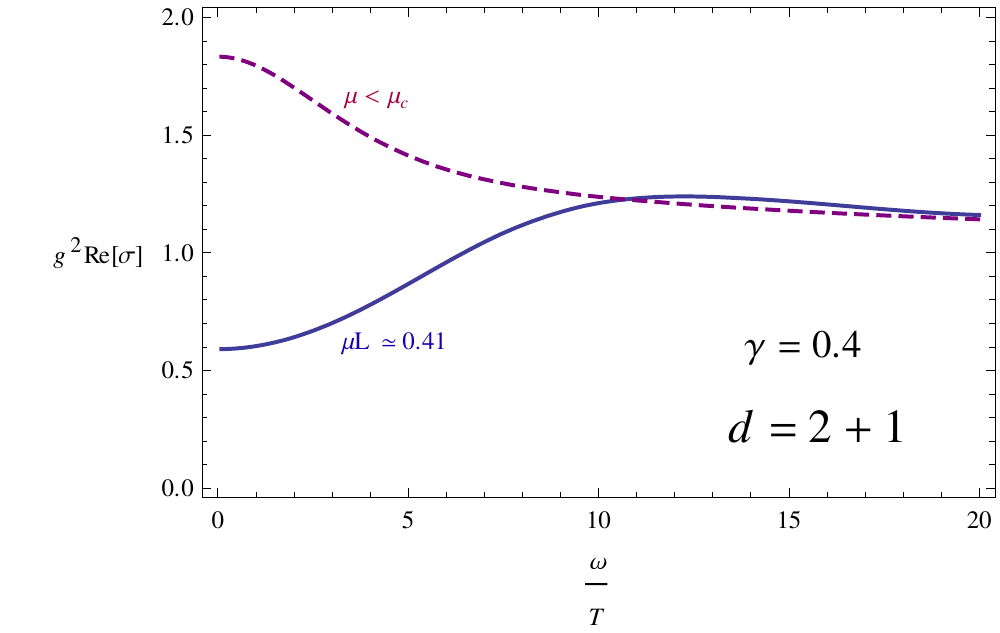}  
      {\hspace{0.2cm}} 
    \includegraphics[scale=0.66]{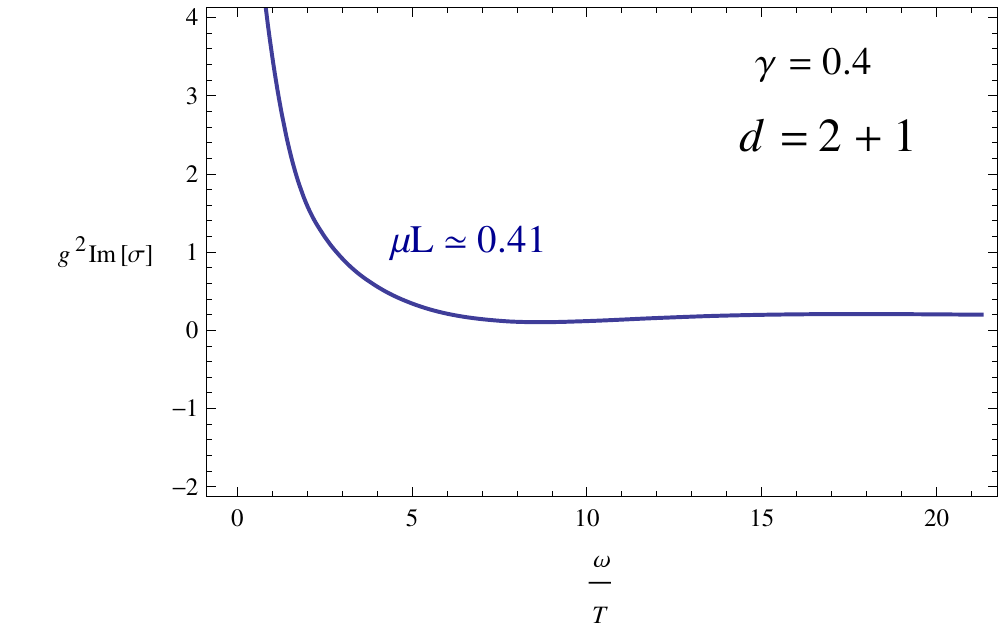}  
    \end{tabular}

    \vspace{0.4cm}
\begin{tabular}{cc}
   {\hspace{-0.7cm}}
   \includegraphics[scale=0.65]{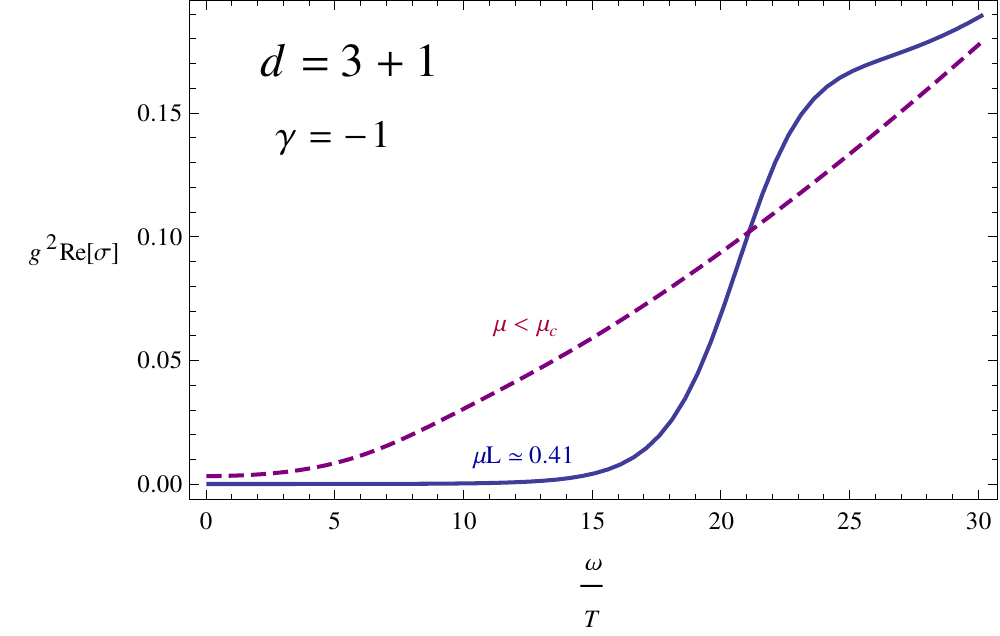}  
      {\hspace{0.3cm}} 
    \includegraphics[scale=0.65]{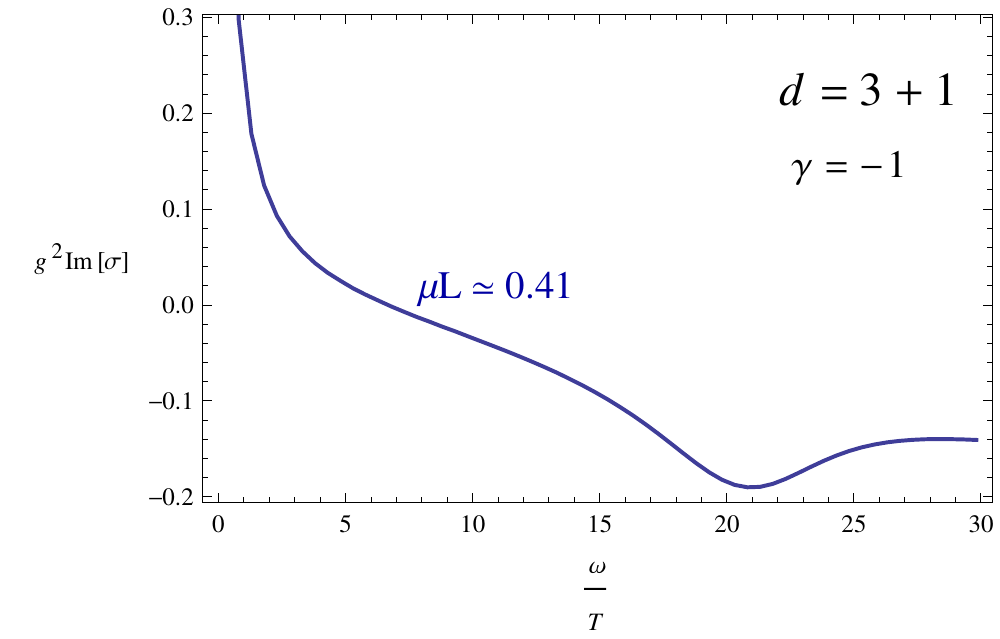}  
    \end{tabular}
    
     \vspace{0.4cm}
     \begin{tabular}{cc}
   {\hspace{-0.7cm}}
   \includegraphics[scale=0.64]{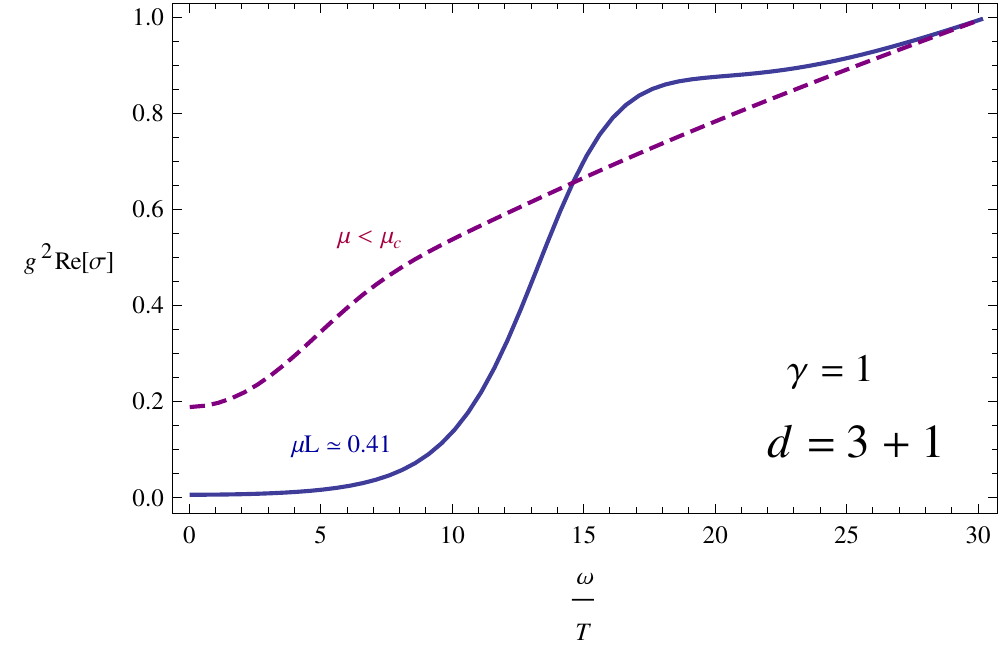}  
      {\hspace{0.2cm}} 
    \includegraphics[scale=0.66]{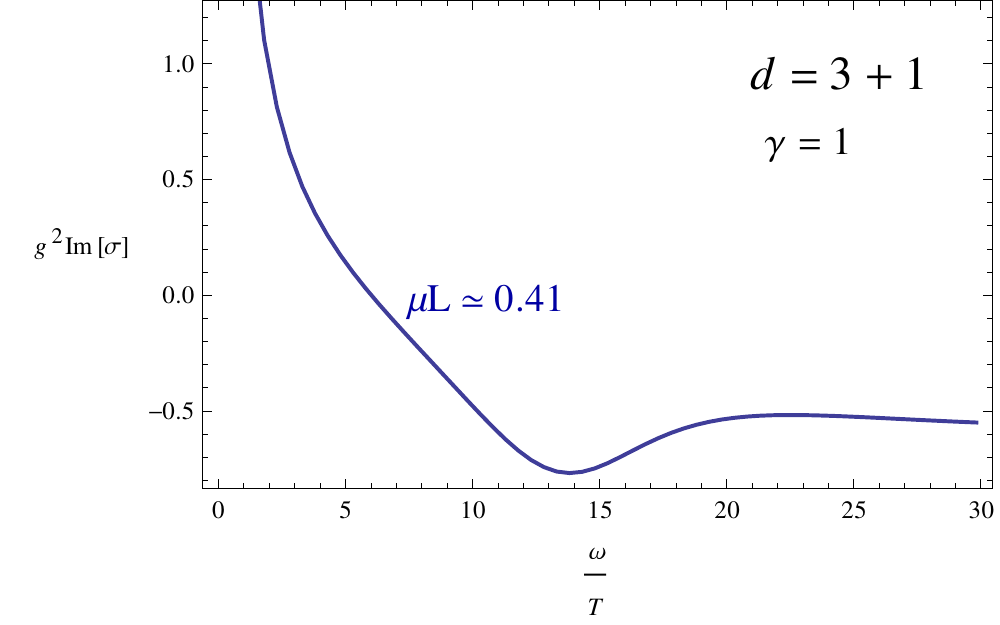}  
    \end{tabular}

   \caption{\label{sigma-SP}The conductivity as a function of the frequency  in the broken phase, $\mu> \mu_c$, (solid blue lines)
   and in the unbroken phase (dashed purple line). Re$[\sigma]$ has also a term  $\propto \delta(\omega)$ in the broken phase. 
   We set $T\simeq 0.010/L$, $\nu=1.2$, $Z_A(\phi)=e^{\gamma\phi}$ and $Z_\psi=1$. 
   Moreover, we choose the bare kinetic term such that  $g^2/e_b^2=L$ ln$(z/L)|_{z=0}$
for $d=3+1$,  
   while $e_b/g\rightarrow \infty$ for $d=2+1$.
}

\end{figure}

\subsection{Conductivity in the broken phase}\label{conductivity-broken}

The broken phase should display an infinite DC conductivity as the system is superconducting.
We provide $\sigma$ as a function of $\omega/T$ in figure \ref{sigma-SP}. As $\omega\rightarrow 0$ the imaginary part diverges like $1/\omega$. This corresponds, through the  Kramers-Kronig relation
\be \mbox{Im}[\sigma(\omega)] = -\frac{1}{\pi} \mathcal{P} \int_{-\infty}^\infty d\omega'\frac{\mbox{Re}[\sigma(\omega')]}{\omega' -\omega }\ ,\ee
to a delta function in the real part: Re$[\sigma(\omega)]\sim \pi n_s \delta(\omega)$. This indeed means that the system 
is superconducting (a superfluid in the fluid mechanical case).  The quantity $n_s$ can be identified with the superfluid density
and from figure \ref{sigma-SP} we see that  $n_s$ is generically changed by changing the couplings  of the dilaton.

Moreover, we  see that   the condensate leads to a  gap $\omega_g$ for charged excitations in the plot of Re$[\sigma]$ versus $\omega$, 
like in the S-BH  configuration \cite{Hartnoll:2008vx,Horowitz:2008bn}. Following \cite{Horowitz:2008bn}, we identify $\omega_g$ with the frequency
at which Im$[\sigma]$ has a bump (in figure \ref{sigma-SP} a minimum).
We see that in the dilaton-BH, the gap depends on the couplings of the dilaton. We have also computed the ratio $\omega_g/T_c$ to compare it with 
the BCS value, around $ 3.5$, and with the result for the S-BH, around $8$ with deviations of less than 8\%, as discussed in \cite{Horowitz:2008bn}.
For the values of the parameters chosen in figure \ref{sigma-SP} we find 
\bea \frac{\omega_g}{T_c}&\simeq&  7.4\,\, (8.7) \qquad \mbox{for} \,\,\, d=2+1 \,\,\, \mbox{and} \,\,\, \gamma=0.4 (-1) \ , \\
   \frac{\omega_g}{T_c}& \simeq & 7.9 \,\,(9.1) \qquad \mbox{for} \,\,\, d=3+1 \,\,\, \mbox{and} \,\,\, \gamma=1 (-1) \ ; \eea
therefore changing the couplings of the dilaton changes, although perhaps weakly, this ratio as well. The possibility of having such 
a high $\omega_g/T_c$ compared to the BCS  3.5 is a nice prediction of holographic superconductor models: high-$T$ superconducting
materials show indeed a value which is systematically bigger than 3.5.
We observe that a negative and large value of $\gamma$ corresponds to a bigger value of $\omega_g$ (compared to that obtained
with a positive $\gamma$). It is interesting to observe that a large $\omega_g$ is associated with an insulating normal phase
according to the results of section \ref{conductivity-unbroken}.

\section{Inhomogeneous phases with broken U(1) symmetry}\label{Inhomogeneous phases with broken U(1) symmetry}

All we have discussed so far can be applied both to superconductors and superfluids because 
the dynamics of the gauge field was negligible: more precisely, since we are considering physical phenomena
described by $a_\mu$ and $\mathcal{O}$ only, the backreaction of $\mathcal{O}$ on $a_\mu$ was not important\footnote{However, from
this statement we cannot conclude anything about the interaction between $a_\mu$ and the condensate constituents, whose operators 
we have not introduced.}. In this section, instead, we study inhomogeneous phases where the difference between 
a global and a local U(1) becomes crucial. In table \ref{table} we provide a dictionary between superconductor observables and the corresponding 
quantities in superfluidity, which are relevant for the following discussion. 
Given their importance in superconductivity and superfluidity, we study in particular vortex configurations.

\begin{table}[t]
\begin{center}
\begin{tabular}{|l|l|l|}
\hline {\bf Generic quantity} & {\color{Sepia}\bf  Superfluid (SF)}  & {\color{ Sepia}\bf Superconductor (SC)}  \\ 
\hline
  $\langle J^i \rangle$  & SF current density  &  EM current density  \\
  arg$(\mathcal{O})$& SF velocity potential in the lab frame  &  condensate's phase  \\
  $a_i$ & external velocity in the lab frame  &  EM vector potential  \\
  \hline
  
 {\bf in vortex phases} & {\color{Sepia}\bf  SF}  & {\color{Sepia}\bf SC} \\
  \hline
   {\it $a_\varphi$-behavior}& $a_\varphi$ is frozen &  $a_\varphi\stackrel{large \, r}{\simeq} n+a_1\sqrt{r}e^{-r/\lambda'}$  \\
    {\it  $\psi_{\rm cl}$-behavior}& $\psi_{\rm cl}\stackrel{large \, r}{\simeq}\psi_\infty\left(1-n^2\tilde{\xi}^2/r^2\right)$ at $B=0$ &  $\psi_{\rm cl}\stackrel{large \, r}{\simeq} \psi_\infty+\frac{\psi_1}{\sqrt{r}}e^{-r/\xi'}  $ \\  
{\it quantization of $B$} & No  &  yes:  $\int dr \, r B =n$ \\
 {\it free energy} & $F_n-F_0\stackrel{large \, R}{\sim} n^2 \ln( R/\xi_\GL)- nBR^2/2$ & 
     finite as $R\rightarrow \infty$  \\ 
 {\it first critical field}& $B_{c1}\stackrel{large \, R}{\simeq}2\ln( R/\xi_\GL)/R^2\stackrel{R \rightarrow \infty}{\rightarrow }0 $  &  $H_{c 1} = e_0^2 (F_1 - F_0)/2 \pi$\\
 {\it  second critical field} & $B_{c2}=1/(2\xi_{\rm \GL}^2)$ & $H_{c2}=1/(2\xi_{\rm \GL}^2)$ 
\\
    \hline

 \end{tabular}
\end{center}\caption{Comparison between superconductors and superfluids.  $\lambda'$ and $\xi'$ are the penetration depth and
coherence length; generically $\lambda'\neq \lambda$ and $\xi'\neq \xi$, where $\lambda$ and $\xi$  are the  inverse masses of 
$a_i$ and $\psi_{\rm cl}\equiv |\langle \mathcal{O}\rangle|$ respectively. $\tilde{\xi}$, $a_1$, $\psi_{\infty}$ and  $\psi_1$
are other constants.  $F_n$ is the free energy per unit of volume $V_{d-3}$ 
of a vortex with winding number $n$ and $\xi_\GL$ is the coherence length of the Ginzburg-Landau theory.
  The external magnetic field $H$ is normalized such that it coincides with the total magnetic field  $B$ at $\Psi=0$. For superfluids we have $H=B$.
See ref. \cite{Domenech:2010nf} for a derivation of these model-independent properties using effective field theory methods.} \label{table}
\end{table}

\subsection{Superfluid vortex phase}

Let us start with superfluids. Even in this subsection we will stick to the superconductor notation in order not to introduce additional symbols; the
reader can however translate to the terminology commonly used in superfluidity by means of table \ref{table}.

The simplest example of vortex is a straight line, corresponding to the  ansatz \cite{Albash:2009ix,Montull:2009fe,Albash:2009iq}
\be\Psi=\psi(z,r)e^{i  n\varphi} \ ,\quad A_0=A_0(z,r)\ , \quad
A_{\varphi}=A_{\varphi}(z,r)  \label{ansatz-vortex}\ ,   \ee
where  $n$ is an integer. The straight vortex line is orthogonal to the plane spanned by the polar coordinates 
$dy^2=dr^2+r^2d\varphi^2$, with $0\leq r \leq R$ and $0\leq \varphi <2\pi$, and passes through $r=0$. We imposed $r\leq R$ because, as shown in
table \ref{table}, the superfluid vortex energy $F_n-F_0$ diverges as $R\rightarrow \infty$. We take $R$ to be large compared to the radius size
of the vortex core ($\tilde{\xi}$ in table \ref{table}).

Being irrotational, a superfluid does not rotate if the external angular velocity $B/2$, which we take to be constant for simplicity, is small enough.
By using table \ref{table} we see that $B$ is related to the gauge field through
\be a_{\varphi}=\frac12 Br^2\ . \label{aphiUV}\ee
The simplification which $B= constant$ leads to is that it allows us to have a finite $\langle J_{\varphi} \rangle$ even for $d>2+1$: we do not 
need to renormalize the divergence in (\ref{log-div}) as $\partial^\nu\mathcal{F}_{\nu\mu}=0$ in that case. So in this subsection we will also take $e_b/g\rightarrow \infty$ 
because we do not need any boundary kinetic term. For superconductors this  will not be possible for $d>2+1$ as the dynamics of the gauge field 
forces $B$ to be nonconstant (and in particular to decay exponentially for large r, table \ref{table}) as it will be discussed in the next subsection.

Increasing $B$ a single vortex appears and if $B$ gets even bigger more and more vortices are created until a triangular 
lattice  \cite{triangular} is reached at $B\simeq B_{c2}$, where $B_{c2}/2$ is the maximal angular velocity above which superfluidity is destroyed.
The phase at $B\simeq B_{c2}$ is model-independent as any model of superfluidity reduces to the Ginzburg-Landau theory 
when the condensate becomes extremely small (which is the case at those high $B$). We therefore 
study the single vortex configuration here as it is energetically favorable when the holographic model is expected to deviate substantially from the 
Ginzburg-Landau predictions.

Ansatz (\ref{ansatz-vortex}) leads to the following field equations
\bea
\frac{\partial_z\left(W^{(d-1)/2} f Z_{\psi}(\phi) \partial_z\psi\right)}{W^{(d-1)/2}fZ_{\psi}(\phi)}+ \frac{\partial_r(r\partial_r \psi)}{rf}+\left(\frac{A_0^2}{f^2} -\frac{(A_{\varphi}-n)^2}{r^2f} \right) \psi 
 &=& 0\ , \nonumber \\
\frac{\partial_z\left(W^{(d-3)/2}f Z_A(\phi)\partial_z A_{\varphi}\right)}{W^{(d-3)/2}f Z_A(\phi)}+\frac{r}{f} \partial_r\left(\frac{1}{r} \partial_r A_\varphi\right) - \frac{2W Z_{\psi}(\phi)  \psi^2}{L^2f Z_A(\phi)}  (A_\varphi-n)  &=& 0\ ,\label{eom-vortices} \\
\frac{\partial_z \left(W^{(d-3)/2}Z_A(\phi)\partial_zA_0\right)}{W^{(d-3)/2}Z_A(\phi)}+
\frac{\partial_r\left( r \partial_r A_0\right)}{rf} - \frac{2 W Z_\psi(\phi)\psi^2}{L^2 \, fZ_A(\phi)} A_0 \,  &=& 0 \ , \nonumber\eea
which have to be solved with appropriate boundary conditions. Our (external) conditions are (\ref{UV}) and (\ref{aphiUV}) at the AdS-boundary,
while 
at large $r$, $r=R$, we impose 
\be \partial_r\psi=0\ ,\ \ \  \partial_r A_0=0\ ,\  \ \  A_{\varphi}=\frac{B}{2}R^2\ . \label{superfluid-large-r} \ee
The first two conditions correspond to the physical requirement that the field configuration goes to the homogeneous superfluid 
state, while the last one  is a simple possibility compatible with eq. (\ref{aphiUV}).
Regularity requires the following boundary conditions at $z=z_h$
 \bea f'(z_h)\partial_z \psi +
\frac{\partial_r(r\partial_r \psi)}{r} -\frac{(A_{\varphi}-n)^2}{r^2}
\psi &=& 0\ ,
 \nonumber \\
 f'(z_h)\partial_z A_{\varphi}+r \partial_r\left(\frac{1}{r} \partial_r A_\varphi\right) - \frac{2 W Z_\psi(\phi)\psi^2}{L^2 Z_A(\phi)}(A_\varphi-n)  &=& 0\ ,\label{bch}\\
A_0&=&0\ ,\nonumber\eea 
and at $r=0$

\begin{figure}[ht]

\begin{tabular}{cc}
  \includegraphics[scale=0.67]{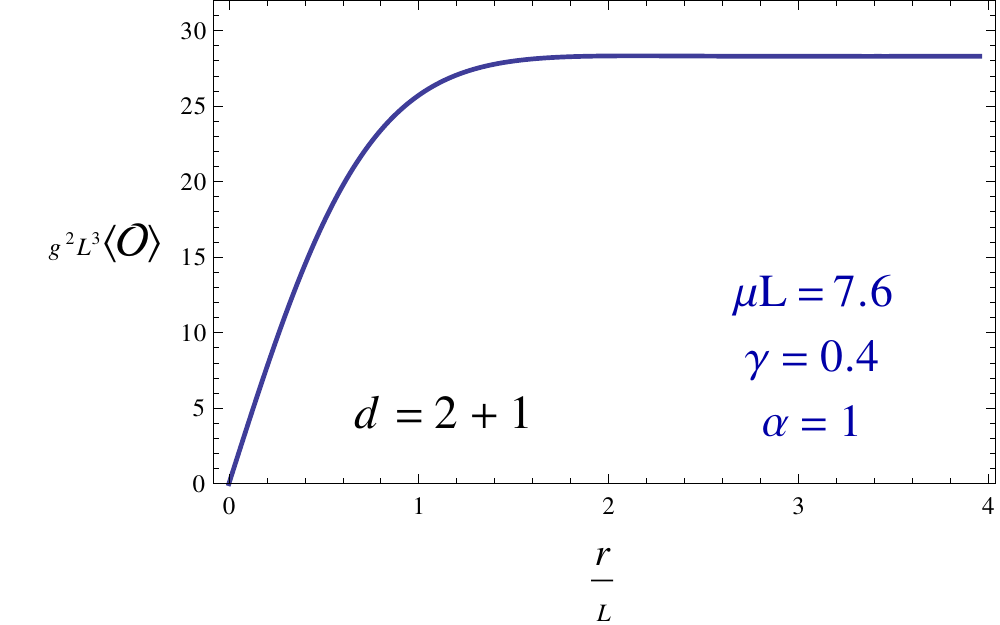}  
 
    {\hspace{0.4cm}} 
    \includegraphics[scale=0.68]{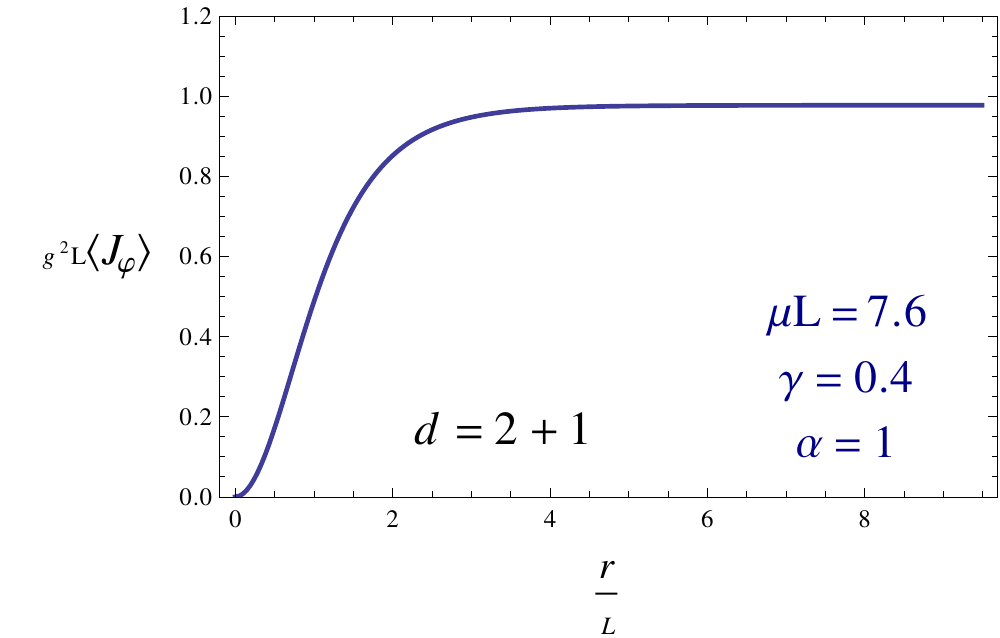}  
   \end{tabular}
   
   \vspace{0.4cm}
   \begin{tabular}{cc}
  \includegraphics[scale=0.68]{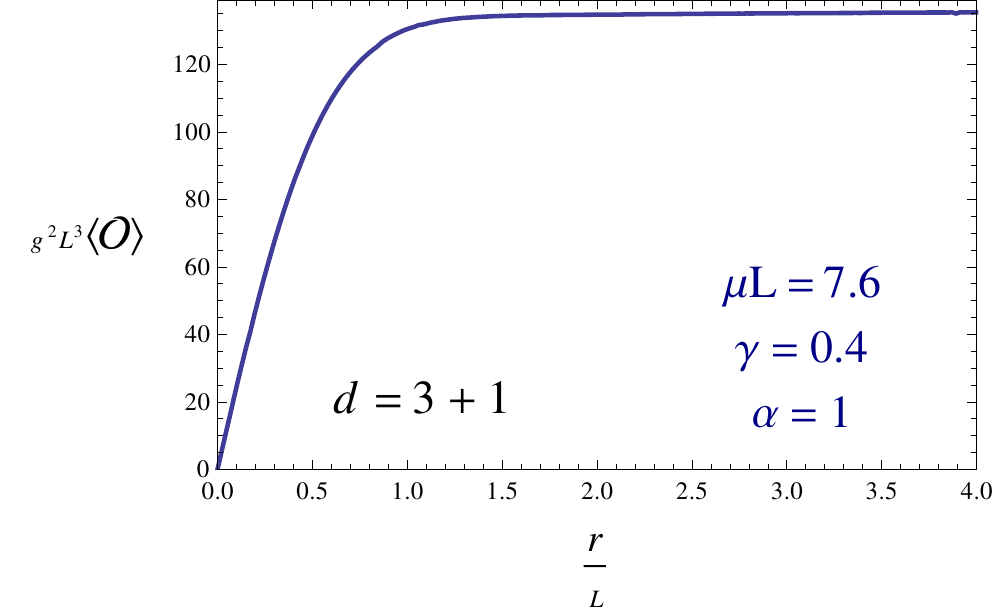}  
 
    {\hspace{0.4cm}} 
    \includegraphics[scale=0.68]{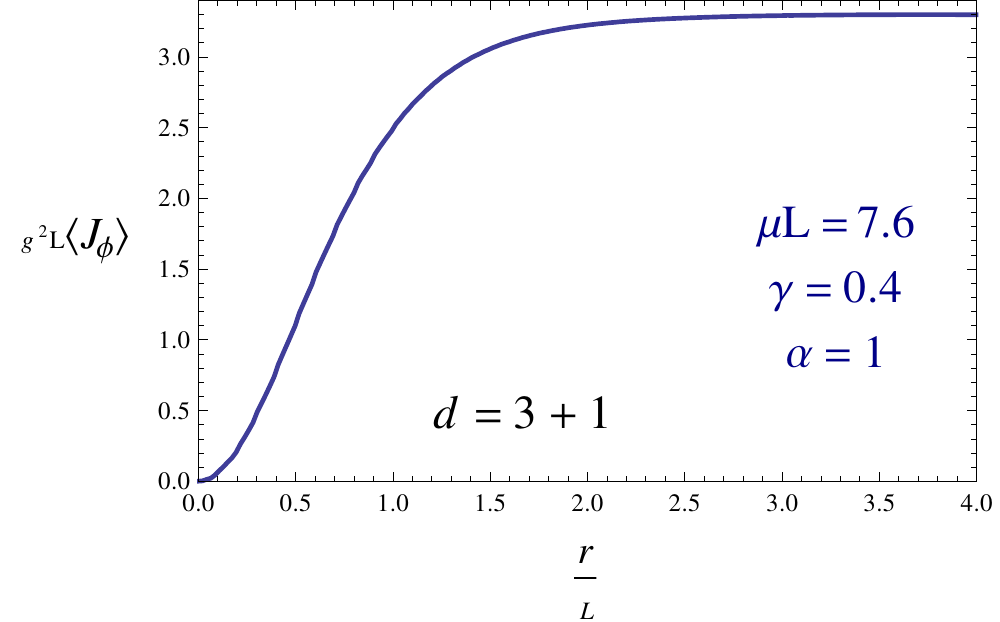}  
   \end{tabular}
   
   \caption{{\small {\it The condensate (left) and the current (right) for the $n=1$ superfluid vortex. 
   In this plot we set $\nu=1.2$, $T\simeq 0.15/L$, $B\simeq 0$, $Z_A(\phi)=e^{\gamma\,\phi}$, $Z_\psi(\phi)=e^{\alpha\phi}$.}}}
\label{OJvsr}
\end{figure}

\bea 
 \partial_r A_0&=&0\ ,\qquad   A_{\varphi} =0\ , \qquad 
 \left\{ \bac  \partial_r \psi=0 \ \  \ \mbox{for}\ n=0\ ,\\ \\ \  \psi=0 \ \ \ \mbox{for}\ n\neq 0\ .
\ea \right.  \label{bcr0} \\
 \qquad  
 \nonumber  \eea

We have solved  eqs. (\ref{eom-vortices}) with the boundary conditions in (\ref{UV}), (\ref{aphiUV}),
(\ref{superfluid-large-r}),  (\ref{bch}) and  (\ref{bcr0}).
In figure \ref{OJvsr} we present the corresponding vortex profiles. 

As shown in table \ref{table}, the maximum value of $B$ for which the system is in the vortex phase, coincides with that for superconductors,
that is $B_{c2}\equiv H_{c2}$. 
We have computed $H_{c2}$ as the value of $B$ at which 
$\Psi \rightarrow 0$ and provide it as a function of $\mu L$ in figure \ref{Hc12-vs-mu} of the next subsection 
(dedicated to superconductors).

\subsection{Superconductor vortex phase}\label{Superconductor vortex phase}

In the superconductor case the dynamics of $a_\varphi$ is important and to properly take into account we use the method described in section 
\ref{The gauge/gravity correspondence}. 
In particular for vortices, the Neumann-like condition at the AdS-boundary, eq. (\ref{maxwell2}), reduces to 
\be
\frac{L^{d-3}}{g^2}z^{3-d}\partial_z A_\varphi\Big|_{z=0} 
+\frac{1}{e_b^2}r\partial_r\left(\frac{1}{r}\partial_r  A_\varphi\right)\Big|_{z=0}=0\label{newbcsc}\ ,
\ee
where we have set $J^\varphi_{ext}=0$. As already mentioned, the second term in the equation above is important in order to 
absorb UV divergences which occur for $d>2+1$. Therefore, since we want to maintain the space-time dimension general, we have to keep that term.
Moreover, the physical requirement that the field configuration goes to the homogeneous superconductor state tells us that at $r\rightarrow \infty$
\be \partial_r\psi=0\ ,\ \ \  \partial_r A_0=0\ ,\  \ \  A_{\varphi}=n\
\label{rinftbc}.\ee

\begin{figure}[t]

   \begin{tabular}{cc}
    {\hspace{0.1cm}}
  \includegraphics[scale=0.63]{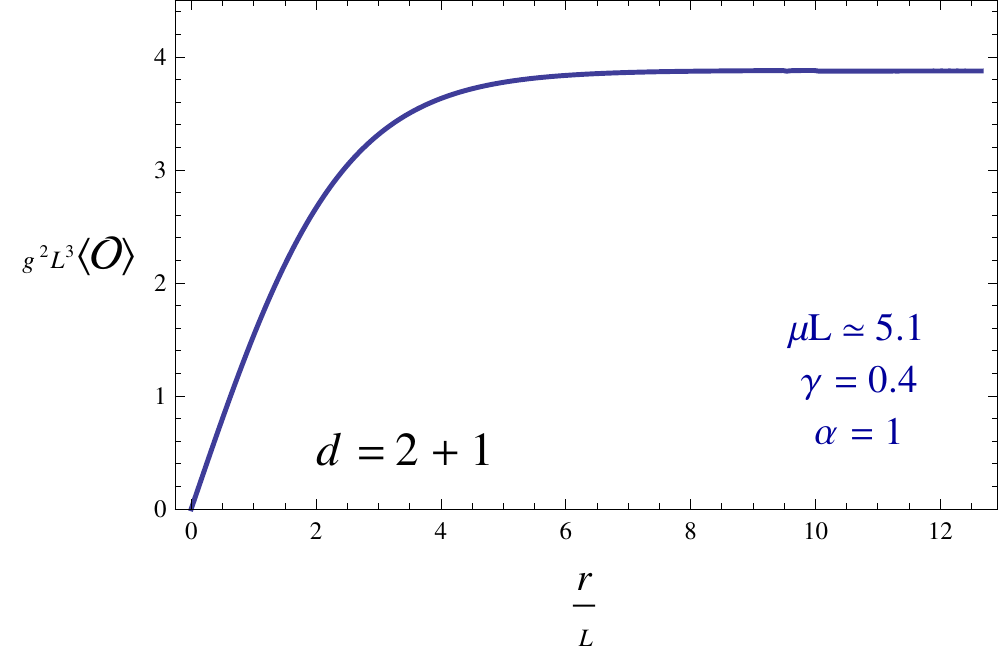}  
    {\hspace{0.4cm}} 
    \includegraphics[scale=0.63]{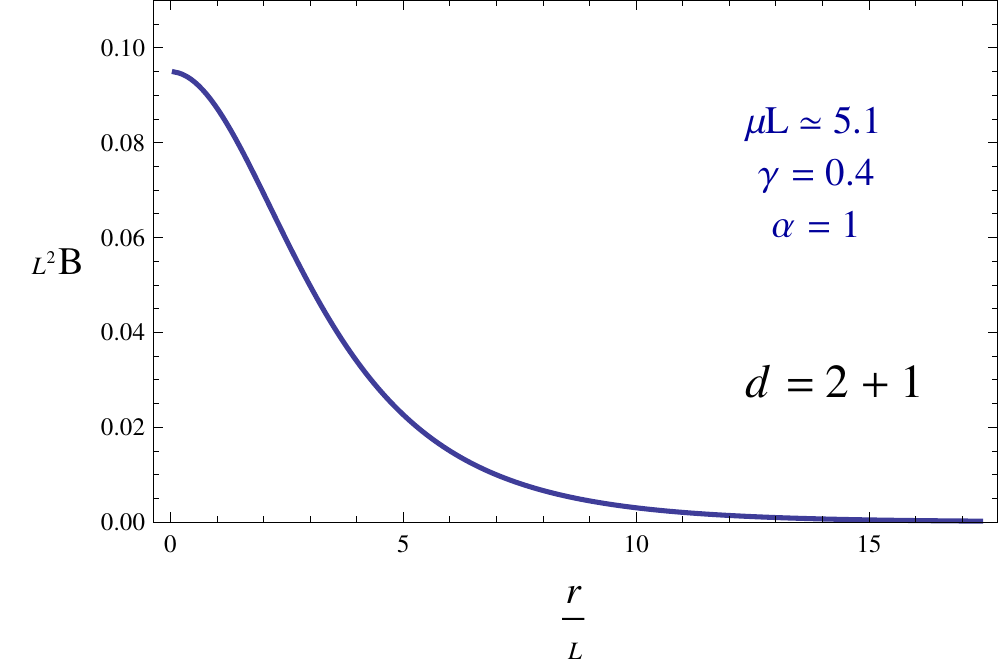}  
   \end{tabular}
   
   \vspace{0.4cm}
 \begin{tabular}{cc}
  \includegraphics[scale=0.67]{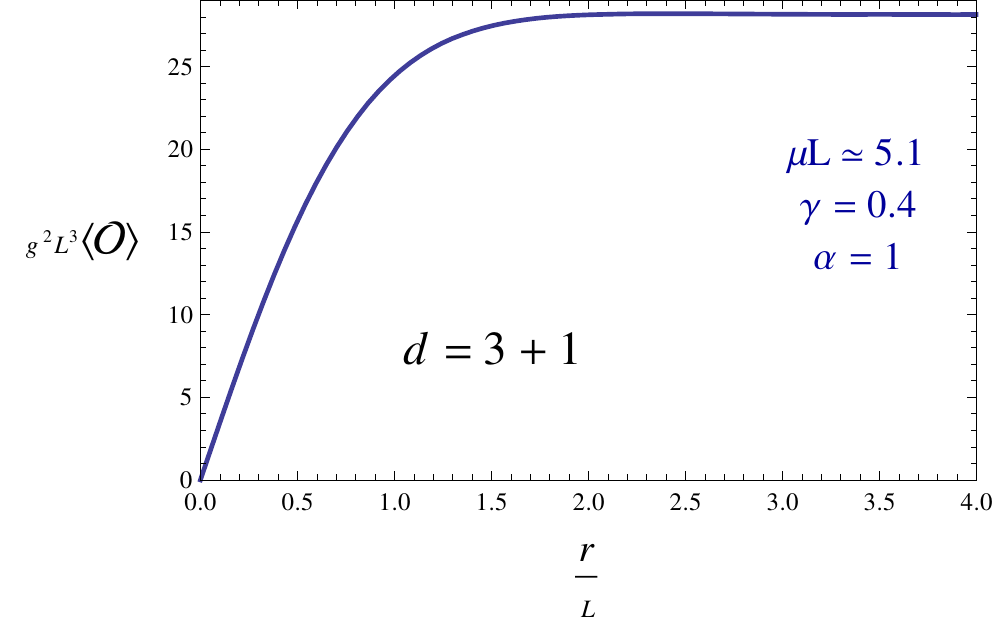}  
 
    {\hspace{0.4cm}} 
    \includegraphics[scale=0.63]{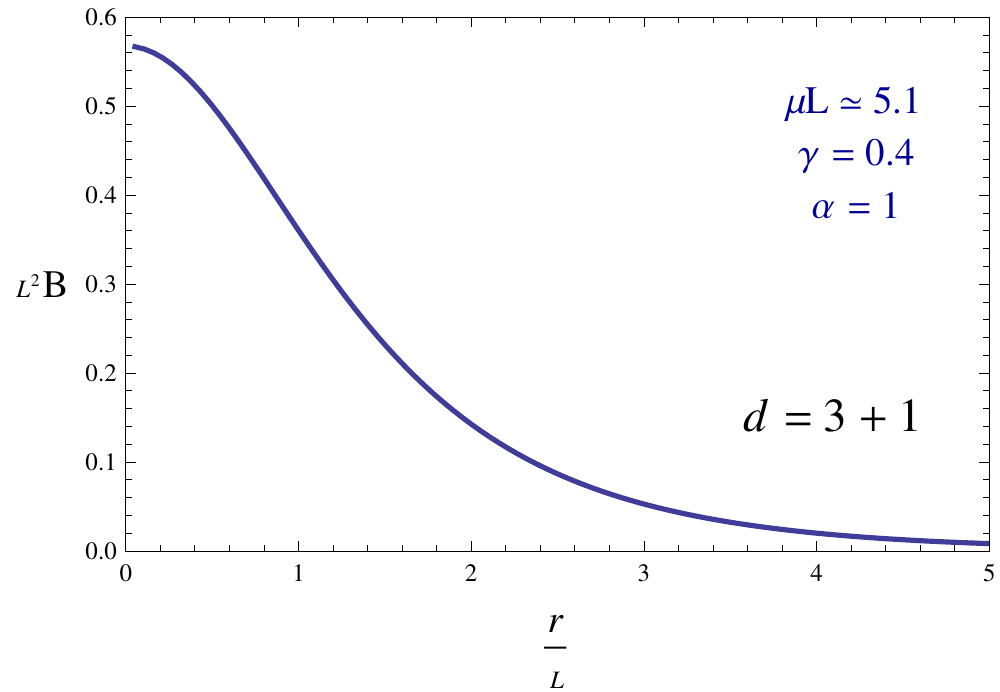}  
   \end{tabular}
   \caption{{\small The condensate (left plots) and the magnetic field (right plots) for the $n=1$ superconductor vortex.
   Here we set $\nu=1.2$, $T\simeq 0.15/L, Z_A(\phi)=e^{\gamma \phi}$, $Z_\psi(\phi)=e^{\alpha \phi}$
and $e_b/g\rightarrow \infty$ for $d=2+1$ while we choose $e_b$ such that $e_0^{-2}\simeq  3.2 L /g^2$ for $d=3+1$.}}
\label{OBvsr}
\end{figure}

\begin{figure}[t]
   \begin{tabular}{cc}
    {\hspace{0.5cm}}
  \includegraphics[scale=0.6]{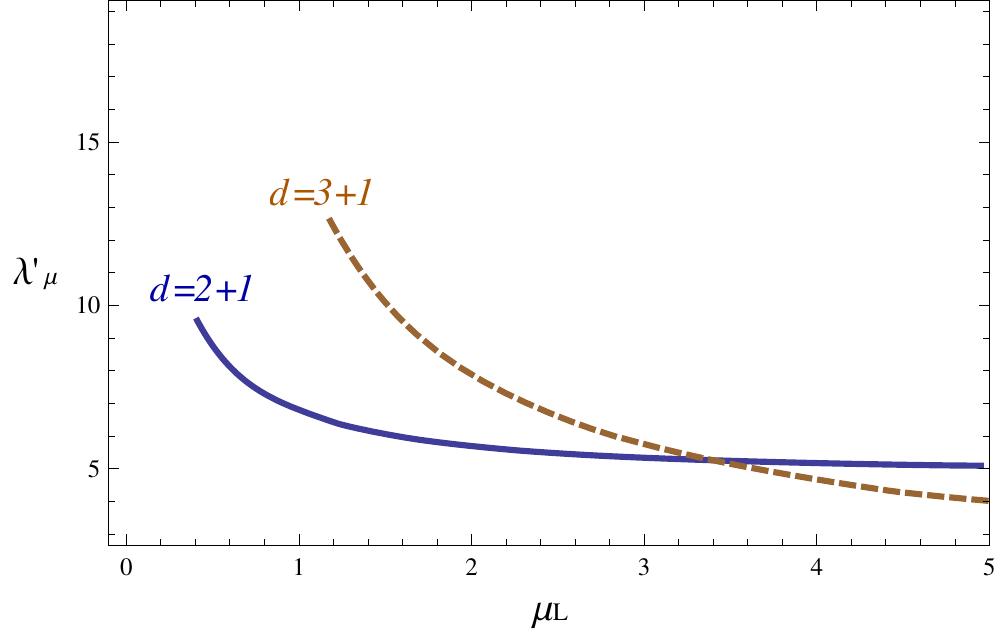}  
 
    {\hspace{1.5cm}} 
    \includegraphics[scale=0.6]{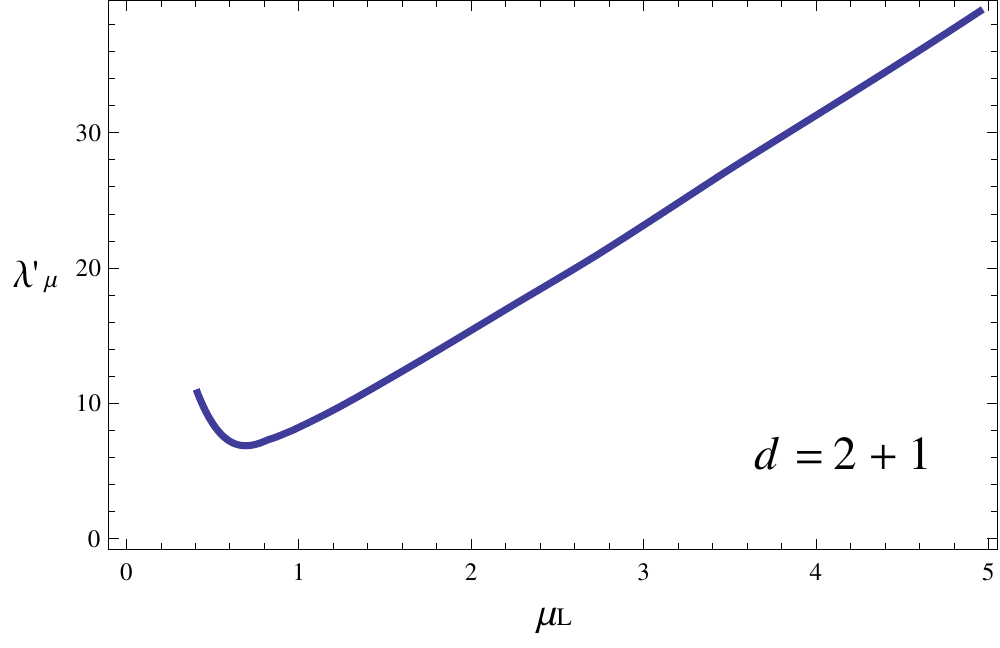}  
   \end{tabular}
   \caption{{\small  The penetration depth as a function of $\mu L$. Here we set $\nu=1.2$, $T\simeq 0.010/L$, $Z_A(\phi)=e^{\gamma \,\phi}$, 
   $Z_\psi(\phi)=1$ and $e_b/g\rightarrow \infty$ for $d=2+1$ while we choose $e_b$ such that $e_0^{-2}\simeq  0.41 L /g^2$ for $d=3+1$.
   We choose $\gamma=-4.4$ on the left and $\gamma=0.4$  on the right.}}
\label{penetration-depth-vs-mu}
\end{figure}

We have solved eqs. (\ref{eom-vortices}) with the boundary conditions in (\ref{UV}), (\ref{newbcsc}),
(\ref{rinftbc}),  (\ref{bch}) and  (\ref{bcr0}).
For $n=0$ the only solution we find is the homogeneous configuration of section \ref{Homogeneous superfluid transitions} with $B=0$: 
this is the celebrated Meissner effect of superconductors.
In figure \ref{OBvsr} we give $\langle \mathcal{O} \rangle$ and $B=\partial_r a_\varphi/r$ as a function of $r$ for the $n=1$ vortex.
Our solutions respect the model-independent properties given in table \ref{table}. This allows us in particular to obtain the coherence 
length $\xi'$ and penetration depth $\lambda'$. We find
\be \xi' \simeq \frac{7 (2) }{\mu}\ , \qquad \mbox{for} \ \ \ d=2+1(3+1) \ee
for the values of the parameters chosen in figure \ref{OBvsr}, while we plot $\lambda'$ as a function of $\mu L$ in figure \ref{penetration-depth-vs-mu}.
 Taking $Z_A(\phi)= e^{\gamma \phi}$ for the sake of definiteness, we observe that when $\gamma$ is negative enough $\lambda'$
 behaves in a qualitatively different way for large $\mu$ with respect to the S-BH. We observe that this is true both for 
 $d=2+1$, as already pointed out in \cite{Salvio:2012at}, and $d=3+1$. Since increasing $\mu L$ for fixed  $TL$ corresponds to exploring 
 the  small $T/T_c$, region we conclude that the penetration depth in the dilaton-BH have a qualitatively different low temperature behavior.
 In particular $\lambda'$ does not increase more and more as  $T/T_c$  decreases,
 unlike in the S-BH \cite{Domenech:2010nf}. 
 We always find that $\lambda'$ diverges at $\mu=\mu_c$: this is a model-independent feature of superconductivity because when 
 $\mu \simeq \mu_c$ the condensate becomes small and
 the Ginzburg-Landau theory holds and predicts that divergence.
 
\begin{figure}[t]
   \begin{tabular}{cc}
    {\hspace{0.5cm}}
  \includegraphics[scale=0.6]{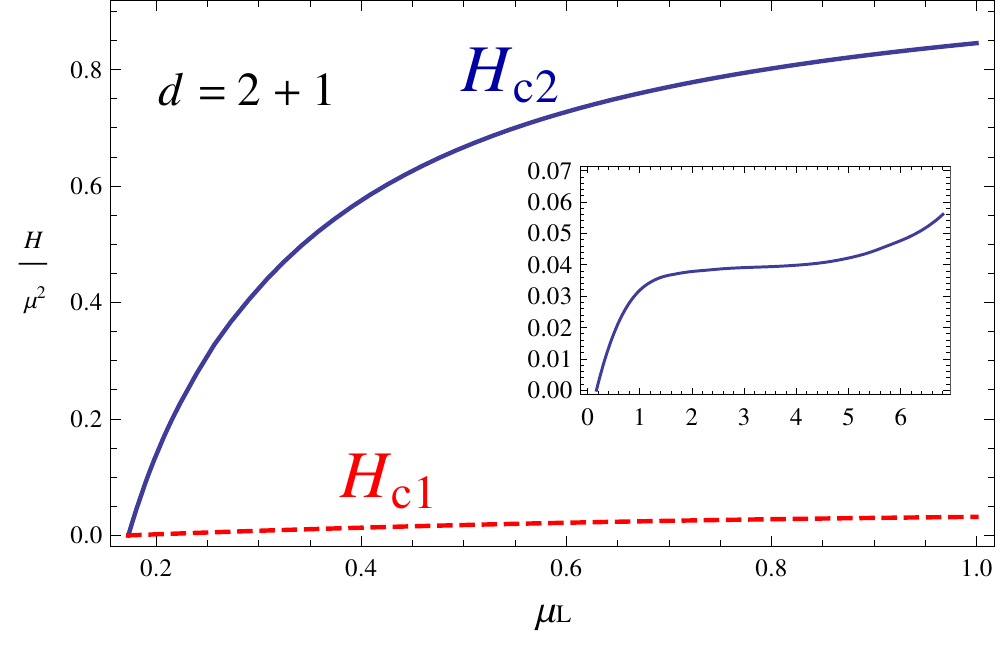}  
 
    {\hspace{1.5cm}} 
    \includegraphics[scale=0.6]{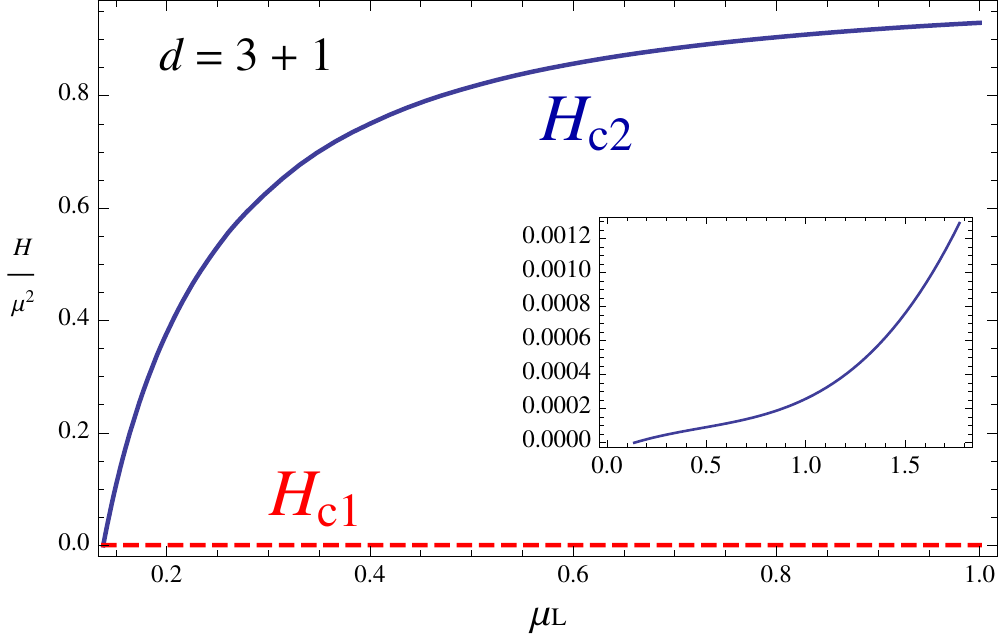}  
   \end{tabular}
   \caption{{\small {\it The critical magnetic fields as a function of $\mu L$ (the inset is an enlarged version of the plot of $H_{c1}$)	.
   Here we set $\nu=1.2$, $T\simeq 0.010/L$, $Z_A(\phi)=e^{\gamma \,\phi}$, $\gamma=-4.4$,  $Z_\psi(\phi)=1$ and $e_b/g\rightarrow \infty$
   for $d=2+1$ while we choose $e_b$ such that $e_0^{-2}\simeq  0.41 L /g^2$ for $d=3+1$.
 }}}
\label{Hc12-vs-mu}
\end{figure}

 Moreover, we have studied the first critical field $H_{c1}$, the value of the external magnetic field $H$ at the boundary between
 the homogeneous superfluid phase and the vortex phase. We can compute this quantity with the text book formula in table \ref{table}.
%
We provide $H_{c1}$ vs. $\mu L$ in figure \ref{Hc12-vs-mu}. Like for the penetration depth, in the high $\mu/\mu_c$ (low $T/T_c$) region we find
substantial qualitative differences compared to the S-BH case if the exponent $\gamma$
 is below a certain bound: while for the S-BH (and for  $\gamma$ above this bound)  $H_{c1}\rightarrow 0$ \cite{Domenech:2010nf}, 
 in the other cases $H_{c1}$ remains sizable even at low $T/T_c$.
This occurs because $H_{c1}\propto e_0^2$, whose low temperature behavior depends strongly on whether  scale invariance is preserved,
$\phi=0$, or broken, $\phi\neq 0$. 
Working out $e_0$ explicitly one finds that in the scale invariant case $e_0 \rightarrow 0$ and so does $H_{c1}$, but when scale invariance
is broken $e_0$ remains finite if $\gamma$ is below a certain bound. From eqs. (\ref{dilaton-BH-AdS4}) and (\ref{dilaton-BH-AdS5})
we see that this bound is $-\sqrt{2/(\nu^2-1)}$ for $d=2+1$ and $\sqrt{(\nu-1)^2/(3(\nu^2-1))}$ for $d=3+1$. Interestingly, we observe that 
in both dimensionalities these values of $\gamma$ lead to an insulating normal phase at low temperatures, but the 
bound for $d=3+1$ is substantially
weaker than that for $d=2+1$.
This conclusion can be generalized to arbitrary
$Z_A(\phi)$ (not necessarily exponential): 
what actually matters here is whether   $Z_A(\phi)|_{z=z_h}$ is small enough for small $T$; but comparing with eq. (\ref{DCsigma}),  
we see that when this occurs the normal phase tends to be insulating.
Finally, the fact that $H_{c1}$ decreases for low $\mu L$  and eventually vanishes at $\mu=\mu_c$ is a model-independent feature of 
superconductivity: at the critical point both the superconductive and the vortex configurations
go to the normal one.

In the same figure \ref{Hc12-vs-mu} we also plot $H_{c2}$, the critical value of $H$ above which superconductivity is completely destroyed. We 
find that $H_{c1}<H_{c2}$ for both $d=2+1$ and $d=3+1$. Therefore, the holographic superconductors under study here are of Type II, like 
their analogues without the dilaton \cite{Domenech:2010nf,Montull:2012fy} and, interestingly, like all known high-$T$ superconductors.
The mere fact that we have Type II superconductors also implies that the phase at high external magnetic fields, $H \simeq H_{c2}$, is a triangular
lattice of vortices \cite{triangular}.

\section{Outlook}\label{conclusions}

In this paper we have investigated various transitions in dilaton holography, 
including in particular those that break a local (superconductor) or global (superfluid) U(1) symmetry in diverse dimensions. 
A detailed summary of the results found is provided in the second part of section \ref{intro}. Therefore, we focus here on possible outlook of the present work.

In this article we have taken the limit in which $G_N\rightarrow 0$ so that the matter fields (the gauge field and the charged scalar) 
do not change the metric and the dilaton. The  next interesting step would be to include the backreaction,
that is going beyond the $G_N\rightarrow 0$ limit. A first step towards this goal could be analyzing the homogeneous 
superfluid phase of section \ref{Homogeneous superfluid transitions} away from such limit. Then, a second (more ambitious) goal might  be finding 
the backreaction of the vortex solutions on the dilaton-gravity system. 
Since a vortex carries angular momentum, one would expect the resulting bulk solution to resemble a rotating charged black hole.

Another interesting extension of this work is the quest for a unified holographic model for the full cuprate phase diagram which includes the insulating (ferromagnetic), superconducting, strange metallic \cite{Hartnoll:2009ns} and pseudogap regions.

\acknowledgments 

I thank Andr\'es Anabal\'on  for useful correspondence and Ioannis Papadimitriou for useful discussions. 
This work has been supported by the Spanish Ministry of Economy and Competitiveness under grant FPA2012-32828, 
Consolider-CPAN (CSD2007-00042), the grant  SEV-2012-0249 of the ``Centro de Excelencia Severo Ochoa'' Programme and the grant  HEPHACOS-S2009/ESP1473 from the C.A. de Madrid.





\end{document}